\documentclass[a4paper,fleqn]{cas-dc}

\usepackage[numbers]{natbib}
\usepackage[official]{eurosym}
\usepackage[shortcuts]{extdash}
\usepackage{amssymb}
\usepackage{multirow}
\usepackage{graphicx}

\begin{document}
\let\WriteBookmarks\relax
\def\floatpagepagefraction{1}
\def\textpagefraction{.001}
\shorttitle{Privacy Preservation in Federated Learning}
\shortauthors{Nguyen Truong et~al.}

\title [mode = title]{Privacy Preservation in Federated Learning: An insightful survey from the GDPR Perspective} 

\author[1]{Nguyen Truong}
\cormark[1]
\ead{n.truong@imperial.ac.uk}
\address[1]{Data Science Institute, South Kensington Campus, Imperial College London, London SW7 2AZ, United Kingdom}

\author[1]{Kai Sun}
\ead{k.sun@imperial.ac.uk}

\author[1]{Siyao Wang}
\ead{s.wang18@imperial.ac.uk}

\author[1]{Florian Guitton}
\ead{f.guitton@imperial.ac.uk}

\author[1, 2]{YiKe Guo}
\address[2]{Department of Computer Science, Hong Kong Baptist University, Kowloon Tong, Hong Kong}
\cormark[2]
\ead{y.guo@imperial.ac.uk}

\cortext[cor1]{Corresponding author}
\cortext[cor2]{Principal corresponding author}

\begin{abstract}
In recent years, along with the blooming of Machine Learning (ML)-based applications and services, ensuring data privacy and security have become a critical obligation. ML-based service providers not only confront with difficulties in collecting and managing data across heterogeneous sources but also challenges of complying with rigorous data protection regulations such as EU/UK General Data Protection Regulation (GDPR). Furthermore, conventional centralised ML approaches have always come with long-standing privacy risks to personal data leakage, misuse, and abuse. Federated learning (FL) has emerged as a prospective solution that facilitates distributed collaborative learning without disclosing original training data. Unfortunately, retaining data and computation on-device as in FL are not sufficient for privacy-guarantee because model parameters exchanged among participants conceal sensitive information that can be exploited in privacy attacks. Consequently, FL-based systems are not naturally compliant with the GDPR. This article is dedicated to surveying of state-of-the-art privacy-preservation techniques in FL in relations with GDPR requirements. Furthermore, insights into the existing challenges are examined along with the prospective approaches following the GDPR regulatory guidelines that FL-based systems shall implement to fully comply with the GDPR.
\end{abstract}

\begin{highlights}

\item Provide a novel systematic analysis on privacy preservation in Federated Learning (FL) taking into account the system architecture, threat models, different types of attack as well as the existing solutions in a centralised FL framework.

\item Conduct a comprehensive survey on privacy-preservation study in centralised FL framework following the structure from the systematic analysis.

\item Provide insightful examination on pros and cons of the existing privacy-preserving techniques as well as prospective solution approaches in order for a FL-based service to comply with the EU/UK General Data Protection Regulation (GDPR).

\end{highlights}

\begin{keywords}
Federated Learning \sep Data Protection Regulation \sep GDPR \sep Personal Data \sep Privacy \sep Privacy Preservation
\end{keywords}

\maketitle

\section{Introduction} \label{INT}
We are now living in a data-driven world where most of applications and services such as health-care and medical services, autonomous cars, and finance applications are based on artificial intelligence (AI) technology with complex data-hungry machine learning (ML) algorithms. AI has been showing advances in every aspect of lives and expected to \textit{"change the world more than anything in the history of mankind. More than electricity.”} \footnote{Dr. Kai-Fu Lee, former vice president at Google, https://www.cnbc.com/2019/01/14/the-oracle-of-ai-these-kinds-of-jobs-will-not-be-replaced-by-robots-.html}. However, the AI technology is yet to reach its full potential, also the realisation of such AI/ML-based applications has been still facing long-standing challenges wherein centralised storage and computation is one of the critical reasons.

In most of the real-world scenarios, data, particularly personal data, is generated and stored in data silos, either end-users' devices or service providers' data centres. Most of conventional ML algorithms are operated in a centralised fashion, requiring training data to be fused in a data server. Essentially, collecting, aggregating and integrating heterogeneous data dispersed over various data sources as well as securely managing and processing the data are non-trivial tasks. The challenges are not only due to transporting high-volume, high-velocity, high-veracity, and heterogeneous data across organisations but also the industry competition, the complicated administrative procedures, and essentially, the data protection regulations and restrictions such as the EU General Data Protection Regulation (GDPR)\footnote{https://gdpr-info.eu/} \cite{horvitz2015data}. In traditional ML algorithms, large-scale data collection and processing at a powerful cloud-based server entails the single-point-of-failure and the risks of severe data breaches. Foremost, centralised data processing and management impose limited transparency and provenance on the system, which could lead to the lack of trust from end-users as well as the difficulty in complying with the GDPR \cite{truong2019gdpr}.

To overcome such challenges, Federated Learning (FL), proposed by Google researchers in 2016, has appeared as a promising solution and attracted attention from both industry and academia \cite{konevcny2016first, konevcny2016second, mcmahan2016federated, mcmahan2017communication}. Generally, FL is a technique to implement an ML algorithm in decentralised collaborative learning settings wherein the algorithm is executed on multiple local datasets stored at isolated data sources (i.e., local nodes) such as smart phones, tablet, PCs, and wearable devices without the need for collecting and processing the training data at a centralised data server. FL allows local nodes to collaboratively train a shared ML model while retaining both training dataset and computation at internal sites \cite{konevcny2016first}. Only results of the training (i.e., parameters) are exchanged at a certain frequency, which requires a central server to coordinate the training process (centralised FL) or utilises a peer-to-peer underlying network infrastructure (i.e., decentralised FL) to aggregate the training results and calculate the global model.

The natural advantage of FL compared to the traditional cloud-centric ML approaches is the ability to reassure data privacy and (presumably) comply with the GDPR because personal data is stored and processed locally, and only model parameters are exchanged. In addition, the processes of parameters updates and aggregation between local nodes and a central coordination server are strengthened by privacy-preserving and cryptography techniques, which enhance data security and privacy \cite{geyer2017differentially, wei2020federated, bonawitz2016practical, bonawitz2017practical, phong2018homomorphic}. The FL capability could potentially inaugurate new opportunities for service providers to implement some sorts of ML algorithms for their applications and services without acquiring clients' personal data, hence naturally complying with data protection regulations like the GDPR. Unfortunately, despite the distributed collaborative learning model of FL empowered by privacy-preserving measures, personal information can be stealthily extracted from local training parameters \cite{aono2017privacy, phong2018homomorphic, zhu2019deep, hitaj2017deep, melis2019exploiting}. As a consequence, FL-based service providers still stay within the regulatory personal data protection framework and are still liable for implementing GDPR-compliant mechanisms when dealing with EU/UK citizens.

In this article, we conduct a survey on existing FL studies with an emphasis on privacy-preserving techniques from the GDPR-compliance perspective. Firstly, we briefly review the challenges on data privacy preservation in conventional centralised ML approaches (\textit{Section 2}) and introduce FL as a potential approach to address the challenges (\textit{Section 3}). Secondly, the state-of-the-art privacy-preserving techniques for centralised FL are described with the analysis of how these solutions can mitigate data security and privacy risks (\textit{Section 4}). Thirdly, we provide an insightful deliberation with potential solution approaches of how an FL system can be implemented in order to comply with the EU/UK GDPR (\textit{Section 5}). Unsolved challenges hindering an FL system from complying with the GDPR are also specified along with the future research directions.
\section{Privacy Preservation and GDPR-Compliance in ML-based Systems} \label{2Challenges}

\subsection{Fundamental Background}
ML is a disruptive technology for designing and building intelligent systems that can automatically learn and improve from experience to accomplish a task without being explicitly programmed. For this purpose, an ML-based system builds up a mathematical model (i.e., model training process) based on a sample set (i.e., training data) whose parameters are to be optimised during this training process. As a result, the system can perform better predictions or decisions on a new, unseen task. Typically, an ML task can be formulated as a mathematical optimisation problem whose goal is to find the extremum of an objective function. Thus, an optimisation method is of paramount importance in any ML-based systems.

\subsubsection{Gradient Descent Algorithm}
One of the most widely used optimisation methods for ML, which is also the core of FL, is gradient descent. It is a first-order iterative optimisation algorithm for finding a local minimum of an objective function $f(\theta)$ parameterised by a set of parameters $\theta \in \mathbb{R}^d$ \cite{ruder2016overview}. Consider a samples set $\mathcal{D}={(x_1, y_1), (x_2, y_2),...,(x_m,y_m)}$, and the objective function $f(\theta)$; a model training process uses the gradient descent method to update each parameter in the opposite direction of the gradient of the objective function $\bigtriangledown f(\theta)$ regarding to the parameters by the following equation:

\begin{equation}
\label{eq:gradient-descent}
    w_j \leftarrow w_j - \eta \bigtriangledown \frac{1}{m} \sum_{i=1}^{m} \mathcal{L}(f(x_i) - y_i)
\end{equation}
where $w_j$ refers to the $j^{th}$ parameter of $\theta$, and $\eta$ refers to the learning rate hyper-parameter, i.e., the size of steps to reach the optimal. $\mathcal{L}$ represents a loss function such as mean-square error (MSE) and cross-entropy loss. The parameters update process using Equation \ref{eq:gradient-descent} is iteratively carried out until either an acceptable local minimum is found or the difference of the loss between two consecutive steps is negligible.

\subsubsection{Gradient Descent Variants}
Generally, there are three gradient descent methods that are categorised based on the amount of training data used in the gradient calculation of the objective function $f(\theta)$ \cite{ruder2016overview}. The first category is \textit{batch gradient descent}, in which the gradients are computed over the entire training dataset $\mathcal{D}$ for one update. The second category is \textit{stochastic gradient descent (SGD)}, that, in contrast to batch gradient descent, randomly selects a sample (or a subset) from $\mathcal{D}$ and performs the parameters update based on the gradient of this sample only (one sample per step, the whole process sweeps through the entire dataset). The third one is \textit{mini-batch gradient descent} in which the dataset is subdivided into mini-batches of $n$ training samples ($n$ is the batch-size); the parameters update is then performed on every mini-batch (single mini-batch per step).

There is a trade-off between the accuracy of parameters update and the efficiency of the computation in each step of gradient descent. Generally, mini-batch gradient descent mitigates the problem of inefficiency in batch gradient descent and gradient oscillation in SGD. However, it introduces the extra hyper-parameter batch-size $n$, which requires expertise and extensive trial and error and sometimes needs to be manually adjusted \cite{keskar2016large}. The gradient descent normally comes along with optimisers, which are techniques for controlling the learning rate $\eta$ logistically and accurately. Such optimisers tie together with the model parameters $\theta$ and the loss function $\mathcal{L}$ in order to adjust the learning rate $\eta$ in response to the output of the loss function. The most common gradient-based optimisers include Momentum \cite{qian1999momentum}, Adam \cite{kingma2014adam}, RMSprop \cite{tieleman2012lecture}, and Adagrad \cite{duchi2011adaptive}.

\subsubsection{Gradient Descent in Distributed Learning}
Although gradient descent-based optimisation methods were successfully engaged in various ML algorithms, they have recently re-gained much attention since the emergence of large-scale distributed learning, including FL \cite{bottou2010large, dean2012large}. In these scenarios, a complex model, e.g., a deep neural network (DNN) with millions of parameters, is trained on a very large dataset across multiple nodes. These nodes are called \textit{compute nodes} and grouped into \textit{clusters}. For efficiency, the calculations in the training process should be parallelised using concurrency methods such as \textit{model parallelism} and \textit{data parallelism} \cite{chen2014big}. Model parallelism distributes an ML model into different computing blocks; available computing nodes are then be assigned to compute some specific blocks only. Model parallelism requires mini-batch data is replicated at computing nodes in a cluster, as well as regular communication and synchronisation among such nodes \cite{dean2012large}. Data parallelism, instead, keeps the completeness of the model on each computing node but partitions the training dataset into smaller equal size shards (also known as \textit{sharding}), which are then distributed to computing nodes in each cluster \cite{ben2019demystifying}. The computing nodes then train the model on their subset as a mini-batch, which is especially effective for SGD variants because most operations over mini-batches are independent in these algorithms. Data parallelism can be found in numerous modern ML frameworks including \textit{TensorFlow}\footnote{https://www.tensorflow.org/} and \textit{Pytorch}\footnote{https://pytorch.org/}. The two parallelism techniques can also be combined (so-called Hybrid parallelism) to intensify the advantages while mitigating the drawbacks of each one; as a result, a hybrid system can achieve better efficiency and scalability \cite{chilimbi2014project}.

The architecture of a distributed learning-based system can be centralised (i.e., \textit{master-slave}) or decentralised (i.e., \textit{ring}) \cite{lian2017can}. In a centralised architecture, slaves (i.e., workers) only compute gradients; a master (i.e., a parameter server) obtains the parameters from all workers and disseminates the latest global parameters back to the workers to be updated in the next training round. This centralised distributed learning requires high-communication cost between workers and a server \cite{dean2012large}. In a ring architecture, there is no centralised server to coordinate the parameter update; instead, each node both locally computes gradients and performs parameter aggregation by communicating with other nodes using a \textit{Gossip} algorithm \cite{ram2009asynchronous, daily2018gossipgrad, koloskova2019decentralized}. The ring architecture requires an efficient asynchronous updates strategy among compute nodes; otherwise, \textit{model consistency} cannot be achieved \cite{lian2018asynchronous, ben2019demystifying}.

Nevertheless, both centralised and decentralised architectures are required to acquire model consistency, particularly when data parallelism is employed. There are numerous strategies to update parameters in order to maintain the consistency of a global model, respected to a synchronisation model among compute nodes. In this regard, Asynchronous Parallel (ASP) \cite{recht2011hogwild, dean2012large}, Bulk Synchronous Parallel (BSP) \cite{gerbessiotis1994direct}, and Stale Synchronous Parallel (SSP) \cite{ho2013more} are the most common approaches to update parameters in a distributed learning system. The BSP and the ASP update parameters once receiving all gradients from a bulk of compute nodes (barrier synchronisation) and from just any node (no synchronisation), respectively. Generally, the BSP is relatively slow due to the stall time of waiting whereas ASP is faster as it does not perform any synchronisation; as a trade-off, the convergence in BSP is guaranteed but uncertain in the ASP \cite{pmlr-v80-zhou18b}. The SSP is as an intermediate solution balancing between the BSP and the ASP that performs relaxed synchronisation. In the SSP, compute nodes continue to the next training iteration only if it is not faster than the slowest node by $\beta$ steps, (i.e., the progress gap between the fastest node and the slowest node is not too large), which guarantees the convergence although the number of iterations might be large. However, as a trade-off, the SSP introduce the $\beta$ hyper-parameter which is non-trivial to be fine-tuned \cite{ho2013more}.

\subsection{Privacy Preserving Techniques in ML}
Generally, privacy preservation techniques for a distributed learning system target two main objectives: \textit{(i)} privacy of the training dataset and \textit{(ii)} privacy of the local model parameters (from an optimisation algorithm such as a gradient descent variant) which are exchanged with other nodes and/or a centralised server \cite{shokri2015privacy}. In this respect, prominent privacy-preserving techniques in ML include data anonymisation \cite{narayanan2008robust}, differential privacy \cite{dwork2006our}, secure multi-party computation (SMC) \cite{yao1986generate}, and homomorphic encryption \cite{gentry2010computing}.

\subsubsection{Data Anonymisation}
Data anonymisation or de-identification is a technique to hide (e.g., hashing) or remove sensitive attributes, such as personally identifiable information (PII), so that a data subject cannot be identified within the modified dataset (i.e., the anonymous dataset) \cite{narayanan2008robust}. As a consequence, data anonymisation has to balance well between privacy-guarantee and utility because hiding or removing information may reduce the utility of the dataset. Furthermore, when combined with auxiliary information from other anonymous datasets, a data subject might be re-identified, subjected to a privacy attack called \textit{linkage attack} \cite{fung2010privacy}. To prevent from linkage attack, numerous techniques have been proposed such as \textit{k-anonymity} \cite{sweeney2002k}, \textit{l-diversity} \cite{machanavajjhala2007diversity}, a \textit{k-anonymity}-based method, and \textit{t-closeness} - a technique built on both \textit{k-anonymity} and \textit{l-diversity} that preserves the distribution of sensitive attributes in a dataset so that it reduces the risk of re-identifying a data subject in a same quasi-identifier group \cite{li2007t}.

Unfortunately, such privacy-preserving techniques cannot defend against linkage attacks whose adversaries possess some knowledge about the sensitive attributes. This deficiency in the \textit{k-anonymity}-based methods calls for different approaches that offer rigorous privacy-guarantee such as \textit{differential privacy}.

\subsubsection{Differential Privacy}
Proposed by Dwork \textit{et al.} in 2006, differential privacy \cite{dwork2006our} is an advanced solution of the perturbation privacy\-/preserving technique in which random noise is added to true outputs using rigorous mathematical measures \cite{fung2010privacy}. As a result, it is statistically indistinguishable between an original aggregate dataset and a differentially additive-noise one. Thus, a single individual cannot be identified as any (statistical) query results to the original dataset is practically the same regardless of the existence of the individual \cite{dwork2006our, dwork2008differential, dwork2014algorithmic}. However, there is a trade-off between privacy-guarantee and utility as adding too much noise and improper randomness will significantly depreciate reliability and usability of the dataset \cite{dwork2008differential, dwork2014algorithmic, fung2010privacy}.

Differential privacy technique has been widely employed in various ML algorithms such as linear and logistic regression \cite{chaudhuri2009privacy}, Support Vector Machine (SVM) \cite{rubinstein2012learning} and deep learning \cite{chaudhuri2011differentially, abadi2016deep}, as well as in ML-based applications such as data mining \cite{friedman2010data} and signal processing with continuous data \cite{sarwate2013signal}.

\subsubsection{Secure Multi-party Computation}
SMC, also known as multi-party computation (MPC) or privacy-preserving computation, was firstly introduced by Yao in 1986 \cite{yao1986generate} and further developed by numerous researchers. Its catalyst is that a function can be collectively computed over a dataset owned by multiple parties using their own inputs (i.e., a subset of the dataset) so that any party learns nothing about others’ data except the outputs \cite{goldreich1998secure, canetti1996adaptively, cramer2000general}. Specifically, $n$ parties $P_1, P_2, .., P_n$ own $n$ pieces of private data $X_1, X_2, ..., X_n$, respectively to collectively compute a public function $f(X_1,X_2,..,X_n) = (Y_1,Y_2,..,Y_n)$. The only information each party can obtain from the computation is the result $(Y_1,Y_2,..,Y_n)$ and its own inputs $X_i$. Classical secret sharing such as Shamir’s secret sharing \cite{shamir1979share, brickell1989some} and verifiable secret sharing (VSS) schemes \cite{chor1985verifiable} are the groundwork for most of the SMC protocols.

SMC is beneficial to data privacy preservation in distributed learning wherein compute nodes collaboratively perform model training on their local dataset without revealing such dataset to others. Indeed, SMC has been employed in numerous ML algorithms such as secure two-party computation (S2C) in linear regression \cite{du2004privacy}, Iterative Dichotomiser-3 (ID3) decision tree learning algorithm \cite{lindell2000privacy}, and \textit{k-means} clustering algorithm for distributed data mining \cite{jagannathan2005privacy}. However,most of SMC protocols impose non-trivial overheads which require further efficiency improvements with practical deployment.

\subsubsection{Homomorphic Encryption}
Another approach to preserve data privacy and security in ML is to utilise homomorphic encryption techniques, particularly in centralised systems, e.g., cloud servers, wherein data is collected and trained at a server without disclosing the original information. Homomorphic encryption enables the ability to perform computation on an encrypted form of data without the need for the secret key to decrypt the cipher-text \cite{gentry2010computing}. Results of the computation are in encrypted form and can only be decrypted by the requester of the computation. In addition, homomorphic encryption ensures that the decrypted output is the same as the one computed on the original unencrypted dataset.

Depending on encryption schemes and classes of computational operations that can be performed on an encrypted form, homomorphic encryption techniques are divided into different categories such as partial, somewhat (SWHE), and fully homomorphic encryption (FHE)\cite{acar2018survey}. Some classic encryption techniques, including \textit{Rivest–Shamir–Adleman (RSA)}, is SWHE wherein simple addition and multiplication operations can be executed \cite{acar2018survey}. FHE, firstly proposed by Graig \textit{et al.} in \cite{gentry2009fully, gentry2011implementing}, enables any arbitrary operations (thus, enables any desirable functionality) over cipher-text, yielding results in encrypted forms. In FHE, computation on the original data or the cipher-text can be mathematically transferred using a decryption function without any conflicts.

Even though homomorphic encryption offers rigorous privacy-guarantee to individuals as the original data in plaintext has never been disclosed, there is a practical limitation in performing computation over cipher-text due to the tremendous computational overhead. As a consequence, employing homomorphic encryption in large-scale data training remains impractical \cite{gilad2016cryptonets}.

\subsection{The GDPR}
\begin{figure}[!htbp]
\centering
	\includegraphics[width=0.48\textwidth]{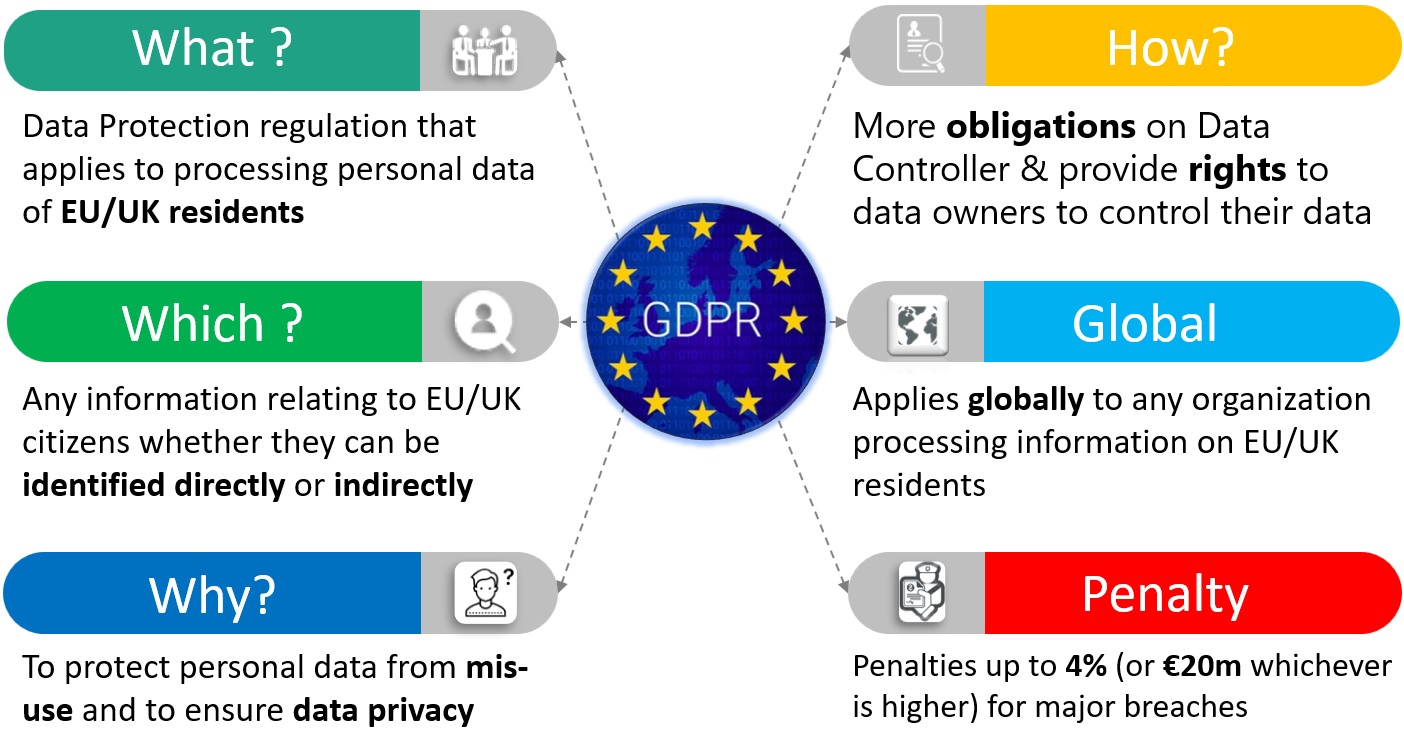}
	\caption{The GDPR legislation in a nutshell}
	\label{fig1}
\end{figure}

The new GDPR legislation has come into force from May 2018 in all European Union (EU) countries which is a major update to the EU Data Protection Directive (95/46/EC) (DPD-95) introduced in the year 1995. The GDPR aims to protect personal data (more comprehensive range depicted in \textit{"Which?"} - Fig. \ref{fig1}) with the impetus that \textit{"personal data can only be gathered legally, under strict conditions, for a legitimate purpose"}. The full regulation is described in detail across $99$ articles covering principles, and both technical and admin requirements around how organisations need to process personal data. The GDPR creates a legal data protection framework throughout the EU/UK member states which has impacted commercial and public organisations worldwide processing EU/UK residents' data (\textit{"Global"} in Fig. \ref{fig1}).

The GDPR clearly differentiates three participant roles, namely: Data Subject, Data Controller and Data Processor, along with associated requirements and obligations under the EU/UK data protection law. While serving as a better privacy and security framework, the GDPR also aims at protecting data ownership by obligating Data Controllers to provide fundamental rights for Data Subjects to control over their data (\textit{"How?"} in Fig. \ref{fig1}). For these purposes, the GDPR introduces and sets high-standard for the consent lawful basis in which Data Controller shall obtain consent from Data Subject in order to process data. Data Controller takes full responsibility to regulate the purposes for which and the methods in which, personal data is processed under the Terms and Conditions defined in the consent.

\subsection{Challenges on Complying with the GDPR}
To meet stringent requirements of the GDPR, conventional ML-based applications and services are required to implement measures that effectively protect and manage personal data adhering to the six data protection principles in the GDPR, as well as to provide mechanisms for data subjects to fully control their data. Although ML-based systems are strengthened by several privacy-preserving methods, implementing these obligations in a centralised ML-based system is non-trivial, sometimes technologically impractical \cite{wachter2017right, greengard2018weighing}.

Large-scale data collection, aggregation and processing at a central server in such ML-based systems not only entail the risks of severe data breaches due to single-point-of-failure but also intensify the lack of transparency, data misuse and data abuse because the service providers are in full control of the whole data lifecycle \cite{truong2019gdpr}. In addition, as ML algorithms operate in a black-box manner, it is also challenging to provide insightful interpretation of how the algorithms execute and how certain decisions are made \cite{mehrabi2019survey, murdoch2019interpretable}. Consequently, most of the ML-based systems find it difficult to satisfy the requirements of transparency, fairness, and automated decision-making in the GDPR.

Furthermore, the requirements of purpose limitation and data minimisation are not always feasibly carried out in ML-based systems. The majority of ML algorithms heavily rely on data quality and quantity, thus researchers tend to collect as much related data as possible. Therefore, determining 1) the purposes of data collection as well as 2) what data is adequate, limited, and relevant only to the claimed purposes before executing such ML algorithms are problematic challenges. These requirements overly restrict the natural operations of ML-based services and applications to a smaller range than ever before.

Finally, ML algorithms are essentially designed for optimising performance, whereas privacy preservation measures remain to be a simple disclaimer. With rigorous requirements of the GDPR, such ML algorithms shall be redesigned internally at the algorithm level in order to accommodate sufficient privacy-preserving techniques. This system redesign requires enormous, or even infeasible, efforts in terms of both technological resolution and human and financial resources. In addition, the trade-off between efficiency and privacy-guarantee is apparently a serious issue for many service providers as sacrificing system performance might lead to the inability to handle their existing services.

\section{Federated Learning: A Distributed Collaborative Learning Approach}
In many scenarios, the traditional cloud-centric ML approaches are no longer suitable due to the challenges of complying with strict data protection regulations on vast aggregation and processing personal data. By nature, most personal data is generated at the edge by end-users' devices (e.g., smart phones, tablets, and wearable devices) which are equipped with increasingly powerful computing capability and Internet connectivity. Given the pervasiveness of such personal devices along with the growing privacy concerns, the trend of decentralised AI has naturally risen which converges the mobile edge computing (MEC) \cite{hu2015mobile} with AI/ML techniques to migrate the intelligence from the cloud to the edge \cite{wang2020convergence}.

In this regard, FL is an alternative for the cloud-centric ML technique that facilitates an ML model to be trained collaboratively while retaining original personal data on their devices, thus potentially mitigates data privacy-related vulnerabilities. It is a cross-disciplinary technique covering multiple computer science aspects including ML, distributed computing, data privacy and security that enables end-users' devices (i.e., local nodes) to locally train a shared ML model on local data. Only parameters in the training process are exchanged for the model aggregation and updates. The difference between FL and the standard distributed learning is that in distributed learning, local training datasets in compute nodes are assumed to be \textit{independent and identically distributed data} (IID) whose their sizes are roughly the same. FL is, thus, as an advancement of distributed learning as it is designed to work with unbalanced and \textit{non-independent identically-distributed data} (non-IID) whose sizes may span several orders of magnitude. Such heterogeneous datasets are resided at a massive number of scattering mobile devices under unstable connectivity and limited communication bandwidth \cite{mcmahan2016federated, mcmahan2017communication, kairouz2019advances}.

\subsection{Model Training in Federated Learning}
FL is well-suited for sorts of ML models that are formulated as minimisation of some objective functions (loss functions) on a training dataset for parameter estimation, particularly for gradient-based optimisation algorithms \cite{konevcny2016first}. The minimisation objective can be formulated as follows:

\begin{equation}
    \min_{w \in \mathbb{R}^d} f(w) = \frac{1}{n} \sum_{i=1}^n f_i(w)
    \label{fl_objective_eq}
\end{equation}
where the training dataset is in form of a set of input-output pairs $(x_i, y_i), x_i \in \mathbb{R}^d$ and $y_i \in \mathbb{R}, \forall i \in \{1,2, .., n\}$. In Equation \ref{fl_objective_eq}, $n$ is the number of samples in the dataset, $w \in \mathbb{R}^d$ is the \textit{parameter vector}, and $f_i(w)$ is a loss function. This formulation covers both linear and logistic regressions, support vector machines, as well as complicated non-convex problems in Artificial Neural Networks (ANN) including Deep Learning \cite{konevcny2016first}. This problem requires an optimisation process that can be efficiently computed by using a gradient descent algorithm with back-propagation technique \cite{rumelhart1985learning, rezende2014stochastic} for minimising the overall loss with respect to each model parameters.

In traditional ML approaches, this sort of algorithms performs a vast number of fast iterations over a large dataset homogeneously partitioned in data servers. Such algorithms require super low-latency and high-throughput connections to the training data \cite{mcmahan2017communication}. Therefore, solving this optimisation problem in the context of FL is different from the traditional ML approaches as such conditions do not hold in FL settings. Training data in FL is unbalanced and non-IID, which is scattered across millions of personal mobile devices with significant higher-latency, lower-throughput connections compared to the traditional techniques working on a cloud-centric data server. In addition, the data and computing resources in personal devices are only intermittently available for training. Therefore, to actualise FL, optimisation algorithms must be well adapted and efficiently performed for federated settings (i.e., federated optimisation \cite{konevcny2016first}).

\subsection{Federated Optimisation}
One of the fundamentals of FL is efficient optimisation algorithms for federated settings wherein training data is non-IID, massively and unevenly distributed across local nodes, first introduced by Kone{\v{c}}n{\`y} \textit{et al.} in 2016 \cite{konevcny2016first}. The distributed settings for the federated optimisation is formulated as follows. Let $K$ be the number of local nodes, $\mathbb{P}_k$ be the set of data samples stored on node $k \in \{1, 2, .., K\}$, and $n_k = |\mathbb{P}_k|$ be the number of data samples stored on node $k$. As personal data in each local node is different, we can assume that $\mathbb{P}_k \cap \mathbb{P}_l = \varnothing$ if $k \ne l$ and $\sum_{k=1}^K n_k = n$. The distributed problem formulation for the minimisation objective is defined as:

\begin{equation}
    \min_{w \in \mathbb{R}^d} f(w) = \frac{n_k}{n} \sum_{k=1}^K F_k(w)
\end{equation}
where the local empirical loss function $F_k(w)$ is defined as:
\begin{equation}
    F_k(w) = \frac{1}{n_k} \sum_{i \in \mathbb{P}_k} f_i(w)
\end{equation}
Here, the $f(w) = \frac{1}{n} \sum_{i=1}^n f_i(w)$ defined in Equation (1) as a convex combination of the local empirical losses $F_k(w)$ available locally to node $k$.

In this federated setting, minimising the number of iterations in the optimisation algorithms is paramount of importance as there is limited communication capability of the local nodes. In the same paper, Kone{\v{c}}n{\`y} \textit{et al.} proposed a novel distributed gradient descent by combining the Stochastic Variance Reduced Gradient (SVRG) algorithm \cite{johnson2013accelerating, konevcny2017semi} with the Distributed Approximate Newton algorithm (DANE) \cite{shamir2014communication} for distributed optimisation called Federated SVRG (FSVRG) \cite{konevcny2016first}. The FSVRG computes gradients based on $\mathbb{P}_k$ data on each local node $k$, obtains a weighted average of the parameters from all the $K$ local nodes, and updates new parameters for each node after round. This algorithm is then experimented based on public Google+ posts, clustered by about $10,000$ users as local nodes, for predicting whether a post will receive any comments. The results show that the FSVRG outperforms the native gradient descent algorithm as it converges to the optimum within only $30$ iterations.

It is worth noting that standard distributed ML algorithms are generally designed to train independent identically\-/distributed (IID) data, and this assumption does not hold in federated settings due to the significant differences of the number of data samples and data distributions among personal mobile devices. Training over non-IID data has been shown to be much less accurate as well as slower convergence than IID data in federated settings \cite{zhao2018federated}. Kone{\v{c}}n{\`y} with his colleagues at Google went further on improving the efficiency of the FSVRG algorithms in distributed settings by minimising the information in parameter update to be sent to an orchestration server \cite{konevcny2016second}. Two types of updates are considered called \textit{structured updates} and \textit{sketched updates} in which the number of variables used in an ML model is minimised as many as possible, along with the compression of the information in the full model updates. Another ambitious federated optimisation approach is that local nodes are independently trained different ML models as a task in a multi-learning objective simultaneously \cite{smith2017federated}. Generally, local nodes generate data under different distributions which naturally fit separate learning models; however, these models are structurally similar resulting in the ability to model the similarity using a multi-tasking learning (MTL) framework. Therefore, this approach improves performance when dealing with non-IID data as well as guarantees the learning convergence \cite{smith2017federated}.

Standing on these federated optimisation research works, McMahan \textit{et al.} proposed a variation of the SGD called \textit{FederatedSGD} along with the \textit{Federated Averaging} algorithm that can train a deep network at $100$ times fewer communications compared to the naive FSVRG \cite{mcmahan2016federated, mcmahan2017communication}. The catalyst of such algorithms is to leverage the increasingly powerful processors in modern personal mobile devices to perform high-quality updates than simply calculating gradient steps. Specifically, each client not only calculates the gradients but also computes the local model for multiple times; the coordination server only performs aggregation of the local models from the clients. This results in fewer training rounds iterations (thus fewer communications) while producing a decent global model. These proposed algorithms well suited for scenarios that are highly limited communication bandwidth with high jitter and latency. In these scenarios, the naive FSVRG algorithms proposed in \cite{konevcny2016first, konevcny2016second} are not efficient enough. Indeed, the algorithms are utilised for a real-world application for text prediction in Google keyboard in Android smartphones (i.e., G-board)\footnote{https://ai.googleblog.com/2017/04/federated-learning-collaborative.html} \cite{yang2018applied}. In this system setting, the \textit{FederatedSGD} is executed locally on the smartphone to compute gradient descent using local data. The gradient is then sent to an aggregation server. This server performs the \textit{FederatedAveraging} algorithm which randomly selects a fraction of smartphones for each training round, and takes the average of all gradients sent from the selected participants to update the global model. This updated global model is distributed to all participants; the local nodes will then update their local models accordingly.

\subsection{Federated Learning Workflow Cycle}
Inspired by the research \cite{bonawitz2016practical, bonawitz2017practical, konevcny2016first, konevcny2016second, mcmahan2016federated, mcmahan2017communication, konevcny2017semi} and the real-world application (i.e., G-board) by the Google team, most of the existing FL-related research works have focused on the centralised FL framework (i.e., centralised FL) wherein an orchestration server plays as a controller requesting and aggregating training results to/from local nodes. However, it does not necessarily require a centralised server for reconstructing a global model; instead, local nodes can directly exchange their training results in a peer-to-peer manner (i.e., decentralised FL) \cite{he2018cola}. This decentralised training approach requires a local updating scheme in which a synchronisation scheme among local nodes must be implemented \cite{ferdinand2020anytime, reisizadeh2019robust} - which is not always feasible in federated settings. Research on decentralised FL is still in its early stage which is either restricted to simple learning models (e.g., linear models) or with the assumption of full or part synchronisation among participants \cite{he2018cola, li2020federated}.

\begin{figure}[!htbp]
\centering
	\includegraphics[width=0.48\textwidth]{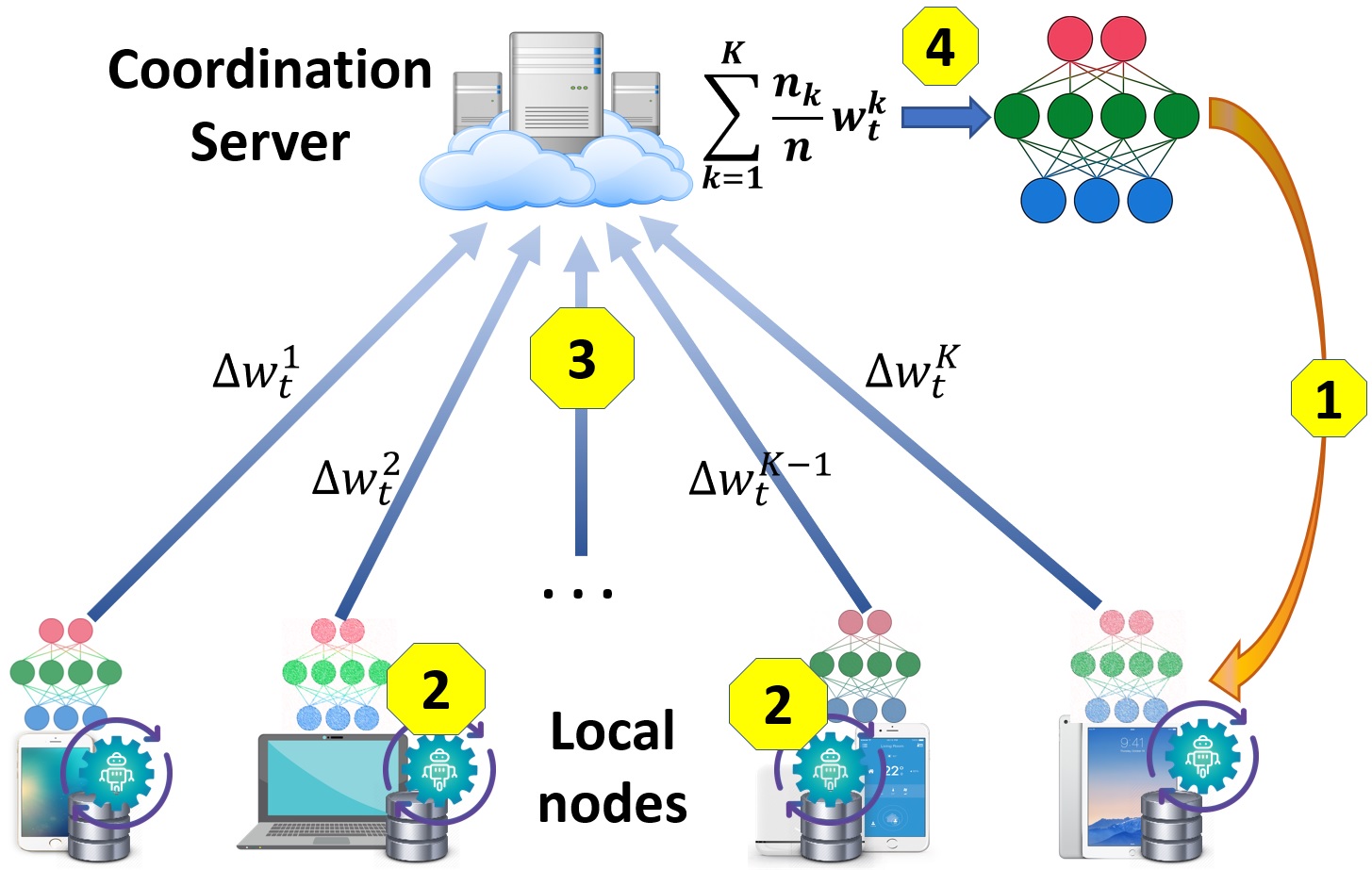}
	\caption{Workflow cycle in a centralised FL framework comprising of four steps}
	\label{fig2}
\end{figure}

In this paper, we examine the centralised FL in which there exists a centralised server (i.e., service provider) requests to coordinate the whole training process. Specifically, this coordination server (i) determines a global model to be trained, (ii) selects participants (i.e., local nodes) for each training round, (iii) aggregates local training results sent by the participants, (iv) updates the global model based on the aggregated results, (v) disseminates the updated model to the participants, and (vi) terminates the training when the global model satisfies some requirements (e.g., accurate enough). Local nodes passively train the model over their local data as requested, and send the training results back to the server whenever possible. The workflow cycle in a centralised FL framework consists of four steps (illustrated in Fig. \ref{fig2}) as follows:

\begin{enumerate}
    \item \textit{Participant Selection and Global Model Dissemination}: The server selects a set of participants that satisfy requirements to be involved in the training process. It then broadcasts a global ML model (or the global model updates) to the participants for the next training round.
    
    \item \textit{Local Computation}: Once receiving the global ML model from the server, the participants updates its current local ML model and then trains the updated model using the local dataset resided in the device. This step is operated at local nodes, and it requires end-users' devices to install an FL client program to perform training algorithms such as \textit{FederatedSGD} and \textit{Federated Averaging}, as well as to receive the global model updates and send the local ML model parameters from/to the server.
    
    \item \textit{Local Models Aggregation}: The server aggregates a sufficient number of the locally trained ML models from participants in order to update the global ML model (the next step). This aggregation mechanism is required to integrate some privacy-preserving techniques such as secure aggregation, differential privacy, and advanced encryption methods to prevent the server from inspecting individual ML model parameters.
    
    \item \textit{Global Model Update}: The server performs an update on the current global ML model based on the aggregated model parameters obtained in step 3. This updated global model will be disseminated to participants in the next training round.
\end{enumerate}
This 4-step cycle is repeated until the global model has reached sufficient accuracy.

It is worth to emphasise that the separation of the four steps in the cycle is not a strict requirement in every training round. For instance, an asynchronous SGD algorithm can be used in which results of the local training can be immediately applied to update the local model before obtaining updates from other participants \cite{chen2016revisiting}. This asynchronous approach is typically utilised in distributed training for deep learning models on a large-scale dataset as it maximises the rate of updates \cite{dean2012large, chilimbi2014project}. However, in FL settings, the synchronous approach, which requires the coordination from a centralised server, has substantial advantages over the asynchronous ones in terms of both communication efficiency and security because it allows advanced technologies to be integrated such as aggregation compression, secure aggregation with SMC, and differential privacy \cite{mcmahan2017communication, konevcny2016second, hardy2017distributed, wang2019adaptive}.
\section{Privacy-Preservation in Centralised Federated Learning Framework}
As an ML model can be cooperatively trained while retaining training data and computation on-device, FL naturally offers privacy-guarantee advantages compared to the traditional ML approaches. Unfortunately, although personal data is not directly sent to a coordination server in its original form, the local ML model parameters still contain sensitive information because some features of the training data samples are inherently encoded into such models \cite{ateniese2015hacking, mcmahan2016federated, aono2017privacy, phong2018homomorphic, melis2019exploiting}. For example, authors in \cite{ateniese2015hacking} have shown that during the training process, correlations implied in the training data are concealed inside the trained models, and personal information can be subsequently extracted. Melis \textit{et al.} have also pointed out that modern deep-learning models conceal internal representations of all kinds of features, and some of them are not related to the task being learned. Such \textit{unintended features} can be exploited to infer some information about the training data samples. FL systems, consequently, is vulnerable to \textit{inference attacks} (i.e., membership and reconstruction attacks \cite{dwork2017exposed}).

Furthermore, local nodes not only passively contribute local training results but also get updated about intermediate stages of a global training model from a coordination server. This practice enables the opportunity for malicious participants to manipulate the training process by providing arbitrary updates in order to poison the global model \cite{fung2018mitigating, bhagoji2019analyzing}, which calls for an investigation on security models along with insightful analysis of privacy guarantees for a centralised FL framework. Accordingly, the FL framework then needs to be strengthened by employing further privacy and security mechanisms to protect personal data effectively and to comply with intricate data protection legislation like the GDPR. A summary of related articles in terms of attack models with associated privacy preservation methods in centralised FL is depicted in Table \ref{tb0}. Detailed descriptions along with analysis are carried out in the following sub-sections.

\subsection{Attack Models on FL}
\subsubsection{Inference Attacks on FL}
As aforementioned, a trained ML model contains unintended features that can be utilised to extract personal information. Thus, local ML model parameters from a federated optimisation algorithm can be exploited by an adversary to infer personal information, particularly when combining with related information such as model data structure and meta-data. This information can be either original training data samples (i.e., \textit{reconstruction attack}) \cite{fredrikson2015model, shokri2015privacy, mcmahan2016federated, aono2017privacy, hitaj2017deep, phong2018homomorphic, shokri2017membership, bagdasaryan2020backdoor, nasr2018comprehensive, zhu2019deep, geiping2020inverting} or \textit{membership tracing} (i.e., to check if a given data point belongs to a training dataset) \cite{bonawitz2017practical, shokri2017membership, melis2019exploiting}.

Attackers might carry out model inversion (MI) attack to extract sensitive information contained in training data samples, for instance, by reconstructing representatives of classes which characterising features in classification ML models \cite{fredrikson2015model}. MI attacks do not require the attacker to actively participate in the training process (i.e., black-box or passive attacks). For example, it is possible to recover images from a facial recognition model for a particular person (i.e., all class members depict this person) using MI by deriving a correct weighted probability estimation for the target feature vectors \cite{shokri2017membership, geiping2020inverting}. In this scenario, the experiment results show that this MI attack can reconstruct images that are visually similar to the victim's photos \cite{fredrikson2015model}.

\begin{figure}[!htbp]
\centering
	\includegraphics[width=0.48\textwidth]{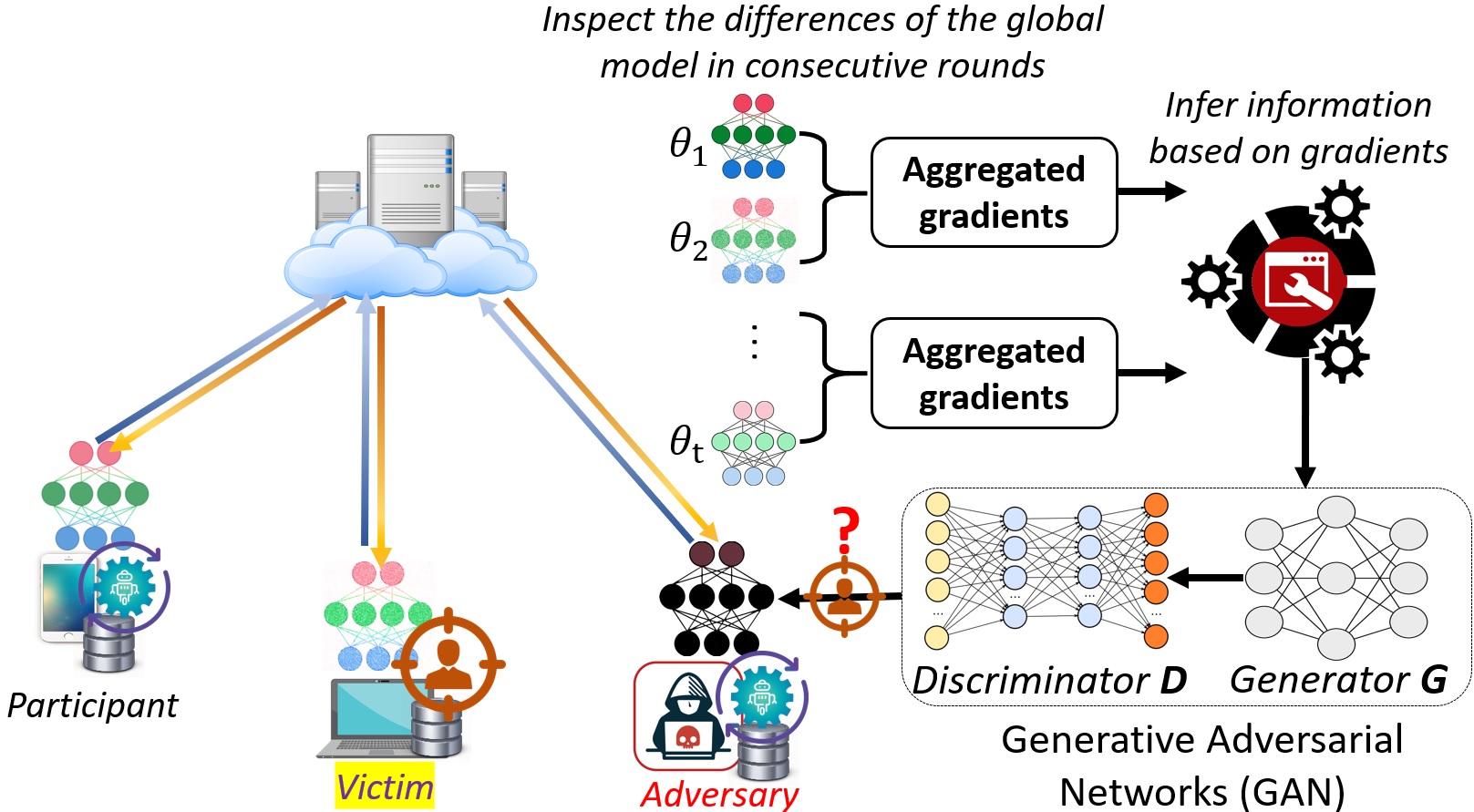}
	\caption{High-level concept of inference attacks against FL based on GANs}
	\label{fig4}
\end{figure}

In FL framework, attackers are not only able to observe the trained model parameters but also participate in the training process to inspect the changes in the updated global models in some consecutive training rounds (i.e., white-box or active attacks), which will intensify the attack (Fig. \ref{fig4}). It is shown that MI attacks based on class representation are more challenging than reconstructing from gradients for classification models \cite{geiping2020inverting}. In this regard, numerous reconstruction attacks were proposed based on Generative Adversarial Networks (GANs) \cite{goodfellow2014generative, salimans2016improved} to synthesise fake samples which have same statistics (e.g., distribution) to those in the training set without having access to the original data. For instance, Hitaj \textit{et al.} based on GANs have developed an attack at user-level which allows an insider to infer information from a victim just by analysing the shared model parameters in some consecutive training rounds \cite{hitaj2017deep}. This attack can be accomplished at client-side without interfering the whole FL procedure, even when the local model parameters are obfuscated using DP technique. A malicious coordination server can also recover partial personal data by inspecting the proportionality between locally trained model parameters sent to the server and the original data samples \cite{aono2017privacy, wang2019beyond}.

Reconstruction attacks using MI and GANs are only feasible if and only if all class members in an ML model are analogous which entails a similarity between the MI/GAN-reconstructed outputs and the training data (e.g., facial recognition of a specific person, or MNIST dataset for handwritten digits\footnote{http://yann.lecun.com/exdb/mnist/} used in \cite{aono2017privacy}). Fortunately, this precondition is less practical in most of the FL scenarios.

However, it is not necessary to fully reconstruct the trained data; instead, inferring attributes or membership of the original trained data from local model parameters can also induce serious privacy leakage \cite{ganju2018property, melis2018inference, melis2019exploiting, nasr2018comprehensive, nasr2019comprehensive} (e.g., an attacker can figure out whether a specific data sample (of a patient) is used to train a model of a disease). This is the baseline for the membership attack. Authors in \cite{melis2018inference, melis2019exploiting, nasr2019comprehensive} have investigated membership attacks in FL and demonstrated the capability of these attacks in both passive and active approaches. For instance, the gender of a victim can be inferred with a very high accuracy of $90\%$ when conducting this attack in a binary gender classifier on the FaceScrub dataset\footnote{http://vintage.winklerbros.net/facescrub.html}. Other features, which are uncorrelated with the main task, can also be inferred such as race and facial appearance (e.g., whether a face photo is wearing glasses) \cite{melis2019exploiting}. Nasr \textit{et al.} proposed an active attack approach called \textit{gradient ascent} by exploiting the privacy vulnerabilities of SGD optimisation algorithms. This attack based on the correlation between the local gradients of the loss and the direction and the amount of changes of model parameters when minimising the loss to fit a model to train data samples in the SGD algorithms. This active membership attack was conducted on the CIFAR100 dataset\footnote{https://www.cs.toronto.edu/~kriz/cifar.html} showing a high accuracy of $74\%$ compared to only $50\%$ in passive attack \cite{nasr2018comprehensive, nasr2019comprehensive}.

\begin{table*}[]
\centering
\caption{Summary of Attack Models vs. Privacy Preservation Methods in centralised FL}
\label{tb0}
\resizebox{\textwidth}{!}{%
    \begin{tabular}{|c|c|c|c|}
    \hline
        \multicolumn{2}{|c|}{\textbf{Attack Models}}
        & \begin{tabular}{@{}c@{}}\textbf{Privacy-preserving Techniques}\\\textbf{employed at Server-side}\end{tabular}
        & \begin{tabular}{@{}c@{}}\textbf{Privacy-preserving Techniques}\\\textbf{employed at Client-side}\end{tabular} \\
    \hline \hline
        \multirow{6}{*}{\textbf{Inference Attacks}}
        & \multirow{3}{*}{
            \begin{tabular}{@{}c@{}c@{}}
            Reconstruction Attacks \\
            \cite{fredrikson2015model, shokri2015privacy, mcmahan2016federated, aono2017privacy, hitaj2017deep, phong2018homomorphic, shokri2017membership, bagdasaryan2020backdoor} \\
            \cite{nasr2018comprehensive, zhu2019deep, geiping2020inverting, goodfellow2014generative, salimans2016improved, wang2019beyond}
            \end{tabular}
        }
        & \multirow{6}{*}{
            \begin{tabular}[c]{@{}c@{}}
                SMC \& Secure Aggregation \\
                \cite{mcmahan2016federated, mcmahan2017learning, mcmahan2017communication, bonawitz2016practical, bonawitz2017practical, bonawitz2019towards} \\
                Homomorphic Encryption \cite{phong2018homomorphic, salem2019utilizing}
            \end{tabular}}
            
        & \multirow{6}{*}{
            \begin{tabular}[c]{@{}c@{}c@{}c@{}}
                SMC \& Secure Aggregation \\
                \cite{pathak2010multiparty, mcmahan2016federated, mcmahan2017learning, mcmahan2017communication, bonawitz2016practical, bonawitz2017practical, bonawitz2019towards}\\
                Homomorphic Encryption \cite{phong2018homomorphic, salem2019utilizing} \\
                Batch-level DP \cite{pathak2010multiparty, shokri2015privacy, abadi2016deep, hitaj2017deep}\\
                User-level DP \cite{pathak2010multiparty, hitaj2017deep, geyer2017differentially, mcmahan2017learning, bhowmick2018protection, sun2020ldp}
            \end{tabular}} \\
        &  &  & \\
        & & & \\
        \cline{2-2} 
        & \multirow{3}{*}{
            \begin{tabular}{@{}c@{}}
            Membership Tracing \\
            \cite{bonawitz2017practical, shokri2017membership, melis2019exploiting, melis2018inference, nasr2019comprehensive, goodfellow2014generative, salimans2016improved, aono2017privacy, wang2019beyond}
            \end{tabular}
        } & & \\
        & & & \\
        & & & \\
        \hline
        
        \multirow{6}{*}{\textbf{Poisoning}} 
        & \multirow{3}{*}{
            \begin{tabular}{@{}c@{}}
            Data Poisoning \\
            \cite{biggio2012poisoning, mei2015using, xiao2015feature, koh2017understanding, chen2017targeted, jagielski2018manipulating}
            \end{tabular}
        }
        & \multirow{6}{*}{
            \begin{tabular}{@{}c@{}c@{}}
            Model Anomaly Detection* \cite{fung2018mitigating, jagielski2018manipulating} \\
            \textit{\textbf{*This solution is not feasible}} \\
            \textit{if Secure Aggregation is employed}
            \end{tabular}
        }
        
        & \multirow{6}{*}{ \textit{\textbf{None}} } \\
        & & & \\
        & & & \\
        \cline{2-2}
        & \multirow{3}{*}{
            \begin{tabular}{@{}c@{}}
            Model Poisoning \\ 
            \cite{blanchard2017machine, mhamdi2018hidden, chen2018draco, fung2018mitigating, bhagoji2019analyzing, bagdasaryan2020backdoor}
            \end{tabular}
        } & & \\
        & & & \\
        & & & \\
        \hline
    \end{tabular}%
}
\label{tab:}
\end{table*}

\subsubsection{Poisoning Attacks on FL}
One of the privacy-preserving objectives of centralised FL is that a coordination server is unable to inspect the data or administer the training process at a local node. This, however, prohibits the transparency of the training process; thus, imposes a new vulnerability of a new type of attack called \textit{model poisoning} \cite{blanchard2017machine, mhamdi2018hidden, chen2018draco, fung2018mitigating, bhagoji2019analyzing, bagdasaryan2020backdoor}. Generally, model poisoning attacks aim at manipulating training process by feeding poisoned local model updates to a coordination server. This type of attack is different from \textit{data poisoning} \cite{biggio2012poisoning, mei2015using, xiao2015feature, koh2017understanding, chen2017targeted, jagielski2018manipulating}, which is less effective in FL settings \cite{bhagoji2019analyzing, bagdasaryan2020backdoor} because the original training data is never shared with a server. Thus, this section is mainly dedicated to analysing the model poisoning attacks in FL.

Generally, model poisoning is conducted at the client-side wherein an adversary controls a fraction of participants for a common adversarial goal, either \textit{(i)} corrupting the global model so that it converges to a \textit{sub-optimal} which is an incompetent, ineffective one (i.e., random attack) \cite{blanchard2017machine, mhamdi2018hidden, chen2018draco}, or \textit{(ii)} replace it to a targeted model (i.e., replacement attack) \cite{bhagoji2019analyzing, bagdasaryan2020backdoor}.

\begin{figure}[!htbp]
\centering
	\includegraphics[width=0.48\textwidth]{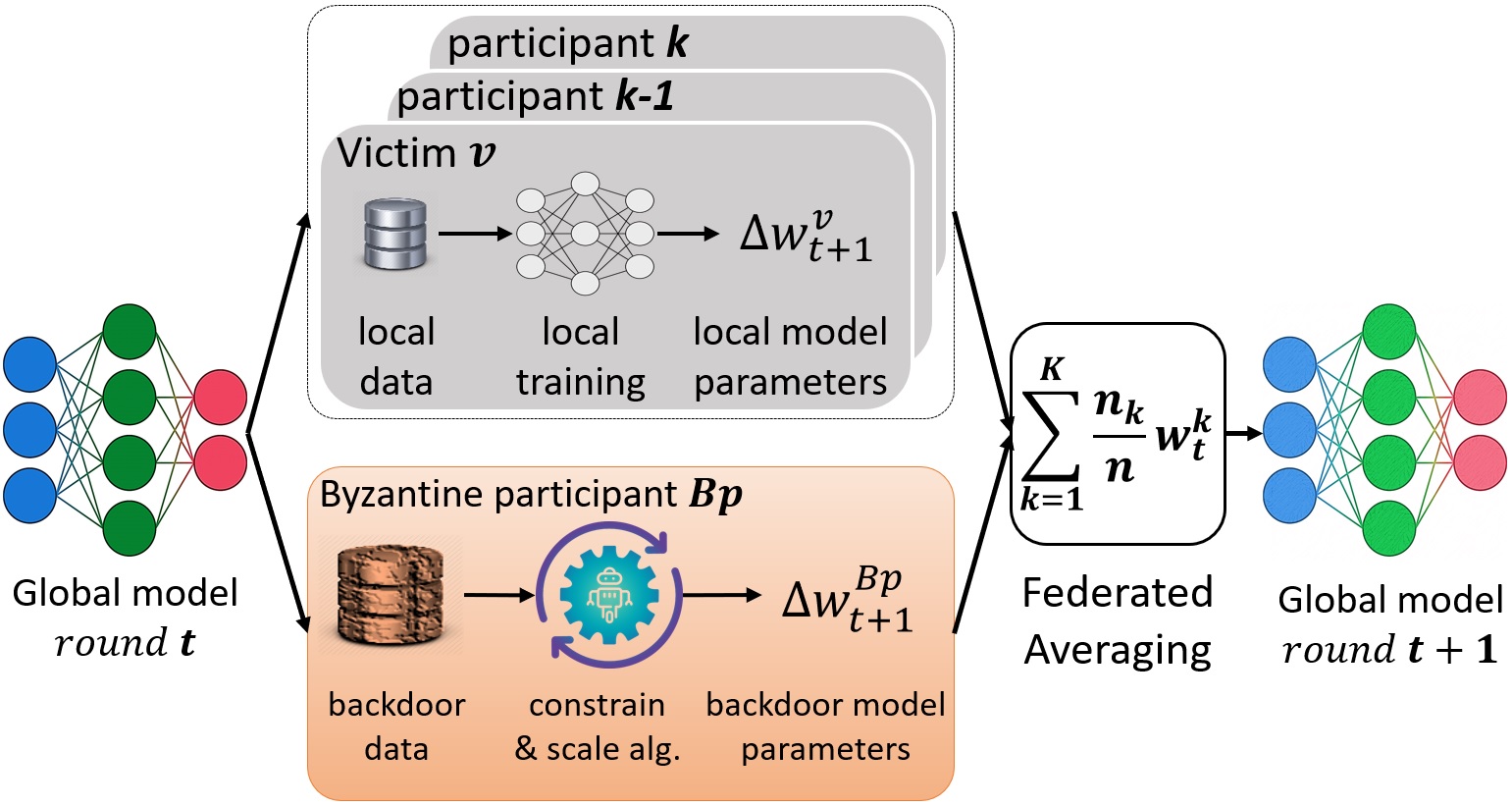}
	\caption{High-level concept of model poisoning using backdoor attack against FL}
	\label{fig5}
\end{figure}

Poisoned model parameters sent to a coordination server can be generated by injecting a hidden backdoor model intentionally, as illustrated in Fig. \ref{fig5}. Compromised participants analyse the targeted global model; the poisoned model is then trained on backdoor data samples using dedicated techniques such as \textit{constrain-and-scale} accordingly, and feed the parameters to a coordination server as other honest participants. The objective of this attack is that the global model is replaced by a joint model consisting of both original task and the injected backdoor sub-task while retaining high accuracy on the two. The backdoor training at the adversary can be empowered by modifying minimisation strategies such as \textit{constrain-and-scale}, which optimises both gradients of the loss and the backdoor objective \cite{bagdasaryan2020backdoor}. A parameter estimation mechanism is then used for generating parameters submitted to the coordination server for honest participants' updates. As secure aggregation is used for preventing the server from inspecting individual models, this poisoning model is unable to detect \cite{bhagoji2019analyzing, bagdasaryan2020backdoor}.

\subsection{Threat model in a centralised FL framework}
\begin{figure}[!htbp]
\centering
	\includegraphics[width=0.4\textwidth]{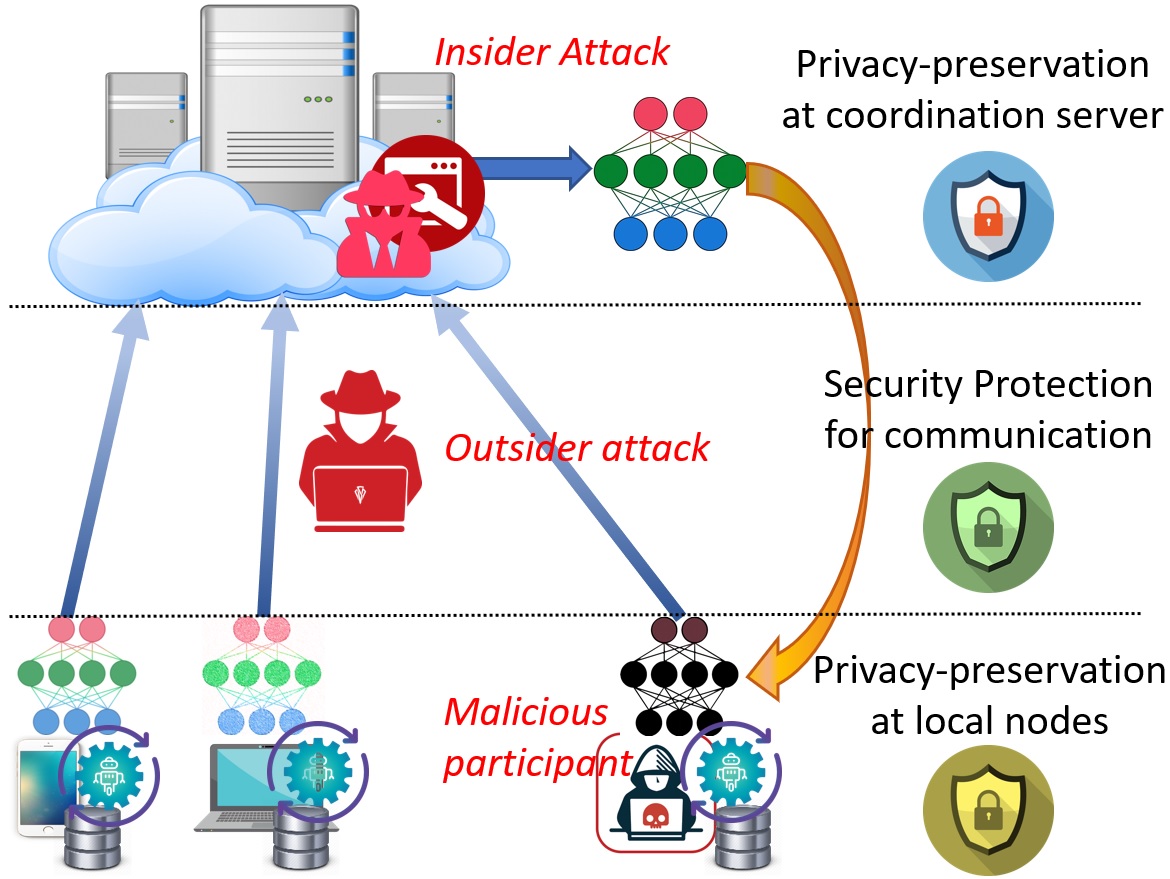}
	\caption{Overview of the Privacy and Security employed in a centralised FL framework}
	\label{fig3}
\end{figure}

As the target of both inference and model poisoning attacks, a centralised FL framework needs to be well designed to withstand potential adversaries. As illustrated in Fig. \ref{fig3}, the security and privacy threats are classified into three categories: (1) Threats at the coordinator server by insider attackers, (2) Threats at communication medium by outsider attacker, and (3) Threats due to malicious participants.

\subsubsection{Malicious coordination server}
The coordination server is assumed to be malicious as there exist insider attackers who can carry out inference attacks to infer information of a target client illegitimately. These attacks are feasible at server-side by analysing periodic parameters updates obtained from related local nodes including the victim (i.e., passive attack), or even purposely requesting the victim to train modifying models with adversarial influence (i.e., active attack) \cite{wang2019beyond}.

\subsubsection{Secure communication medium}
It is assumed that the communication medium for information exchange between local nodes and a coordination server is secure regardless the information is in plain-text \cite{mcmahan2017communication} or encrypted \cite{mohassel2017secureml}. Secure communications protocols such as SSL/TLS and HTTPS are readily integrated into the FL framework to prevent the man-in-the-middle attacks, eavesdropping and tampering. Thus, in a centralised FL framework, privacy and integrity of the exchanged information are assured while in transit.

\subsubsection{Byzantine participants}
In most of FL scenarios, local nodes are assumed to be malicious, meaning that there is a possibility that there exists an adversary controlling a fraction of local nodes to perform \textit{model poisoning}. Moreover, such malicious participants might operate in a Byzantine fashion, meaning that they send arbitrary training model updates to shape the global model in a targeted manner (i.e., either demolish the global model or be replaced by a vicious one).

Furthermore, inference attack can also be carried out a malicious participant as the adversary can commit its local update and observe the changes in the updated global model \cite{melis2019exploiting}. Instead, the active inference attack is only accomplished by a malicious server.

\subsection{Privacy-preservation solutions for coordination server}
Most of existing privacy-preserving techniques for FL systems are built upon advanced cryptographic protocols, including SMC and differential privacy. At server-side, such techniques are employed in order to \textit{(i)} prevent insiders at the server from conducting inference attacks, and \textit{(ii)} prevent Byzantine participants from conducting model poisoning.

\subsubsection{Inference Attacks Prevention}
\begin{figure}[!htbp]
\centering
	\includegraphics[width=0.48\textwidth]{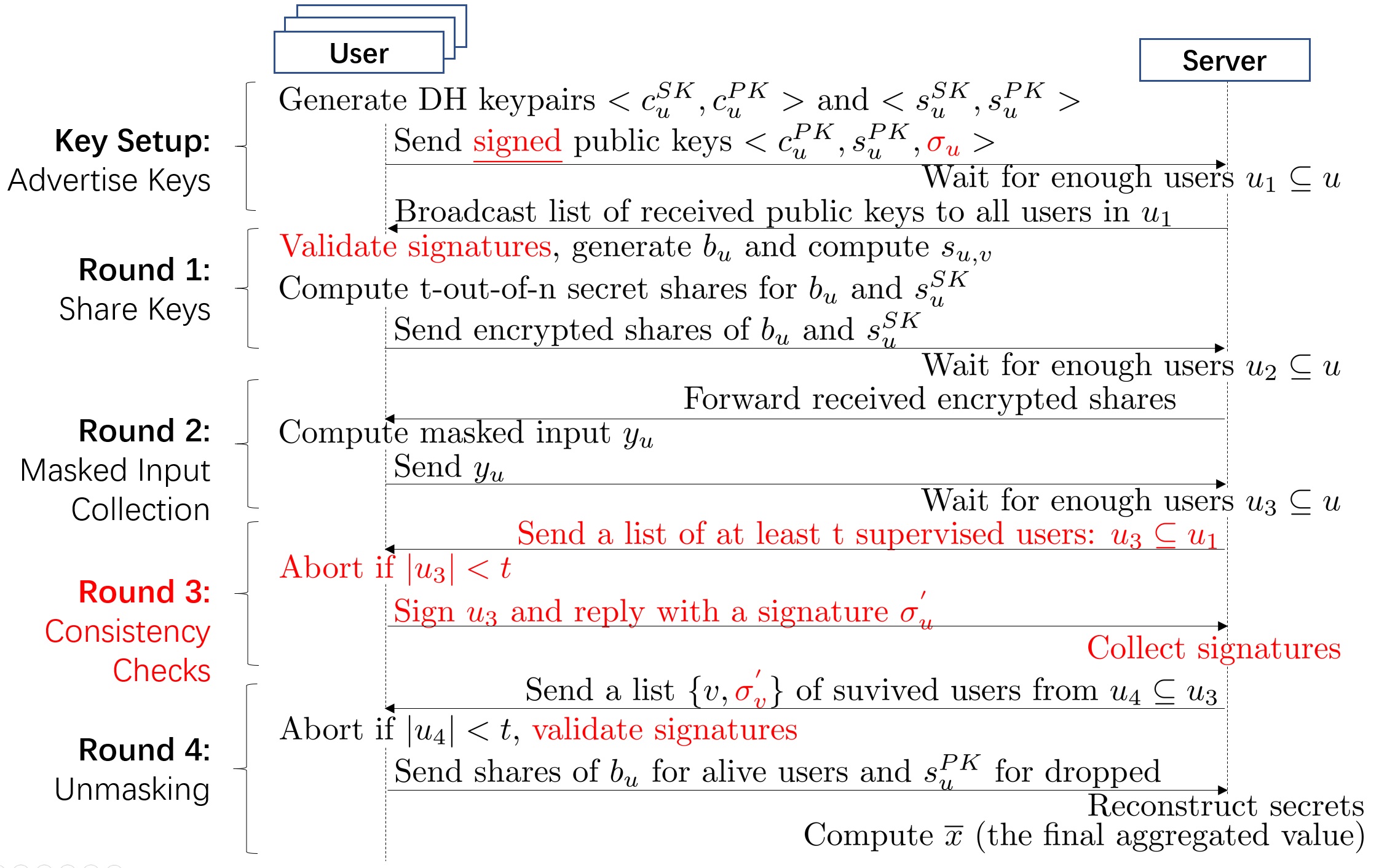}
	\caption{Sequence diagram of the Secure Aggregation. Red-color processes are required to guarantee the security of the protocol against malicious server and participants \cite{bonawitz2017practical}}
	\label{fig6}
\end{figure}

Several solutions have been proposed to tackle against the inference attacks at server-side following the same purpose of preventing the coordinate server from inspecting parameters sent from a particular user during the global model training process. Specifically, in the aggregation process, parameters sent from the clients (\textit{gradients} in Federated SGD or \textit{local model weights} in Federated Averaging) can be protected based on SMC called Secure Aggregation protocol, first proposed by Bonawitz in \cite{bonawitz2016practical, bonawitz2017practical}. The baseline of the protocol is SMC in which cryptography techniques are leveraged that enable participants to jointly compute the average of the model parameters without revealing their inputs. As illustrated in Fig. \ref{fig6}, the protocol comprises of four interactive rounds between participants and a coordinate server including public-keys advertisement and sharing (round 1), masked inputs computation at client-side once getting an independent response from the server (round 2), consistency check that the model has at least $t$ participants involved in the training process (round 3), and unmasking once at least $t$ participants reveal sufficient cryptographic secrets so that the coordination server is able to unmask the global model update (round 4). Round 3 of the protocol is required if the server is malicious but not necessary for an honest-but-curious one. As a trade-off, this protocol results in increasing communication overhead and computation complexity at both clients and a coordination server. It is worth noting that the Secure Aggregation protocol has already been integrated into the TensorFlow Federated framework\footnote{https://www.tensorflow.org/federated}, developed by Google \cite{bonawitz2019towards}, to facilitate research and real-world experimentation with FL.

Secure Aggregation protocol is based on the fact that it only requires to calculate the averages of the local model weights from a random subset of participants to perform SGD and compute global model updates. The coordination server, thus, does not need to acquire local updates from individual participants. This would prevent the server from observing individual users and carrying out inference attacks. Along with Federated Averaging, Secure Aggregation protocol facilitates secure SGD execution with robustness to failures and less communication overhead in a server with limited trust. However, this SMC-based technique only works effectively in scenarios of honest participants. There is no guarantee for the availability and correctness of the protocol in case of Byzantine participants, particularly when such Byzantine participants collude with the malicious server to disclose inputs of a targeted client. In case of the client-server collusion, the protocol can only tolerate up to $[\frac{n}{3}] - 1$ Byzantine participants whereas the number of total participants involved in the training process should be at least $[\frac{2n}{3}] + 1$, ensuring the robustness up to $[\frac{n}{3}] - 1$ dropping out participants \cite{bonawitz2017practical}.

\subsubsection{Model Poisoning Prevention at server-side}
Model poisoning attacks are always inherent in collaborative learning including FL. As shown by Bagdasaryan \textit{et al.} in \cite{bagdasaryan2020backdoor}, just by controlling less than $1\%$ Byzantine participants, an adversary can successfully insert a backdoor functionality into a global model without reducing much accuracy, preventing the coordination server from detecting the attack. Solutions to mitigate model poisoning attack at server-side have to detect and filter out poisoned model updates from malicious clients (i.e., model anomaly detection) \cite{fung2018mitigating, jagielski2018manipulating}. For this purpose, the server needs to access either participants’ training data or parameter model updates, which breaks the privacy-preservation catalyst of FL. Besides, Secure Aggregation protocol is assumed to be implemented at both client- and server-side, which prevents the server from inspecting individual model updates; consequently, ruling out any solutions for model poisoning attacks \cite{fung2018mitigating}. Indeed, no resolutions have been proposed that effectively tackle model poisoning attacks at server-side, which imposes as a critical research topic for FL.

\subsection{Privacy-preservation solutions for local nodes}
Local nodes, along with a coordination server, should implement Secure Aggregation protocol to mitigate the risk of privacy leakage in case there exists an inside attacker carrying out inference attacks at the server \cite{bonawitz2016practical, bonawitz2017practical}. This SMC-based aggregation protocol can also be strengthened with Homomorphic Encryption to encrypt local model parameters from all participants for secure multi-party deep learning in FL settings \cite{zhang2017private}. The coordination server, hence, receives an encrypted global model which can only be decrypted if and only if a sufficient number of local models have been aggregated. As a result, the privacy of individual contributions to the global model is guaranteed.

Furthermore, the local nodes can leverage the perturbation method to prevent a coordination server and other adversaries from disclosing model parameters updates and original training dataset. The idea of employing perturbation technique to FL is that a local node adds random noise to its local model parameters in order to obscure certain sensitive attributes of the model before sharing. As a result, adversaries, in case it can successfully derive such model parameters, is unable to accurately reconstruct the original training data or infer some related information. In other words, the perturbation method could prevent adversaries from carrying out inference attacks on a local model trained by a particular client. This privacy-preservation method typically adopts differential privacy technique that adds random noises to either training dataset or model parameters, offering statistical privacy guarantees for individual data \cite{dwork2014algorithmic, dwork2008differential, bassily2014private}. Indeed, before the proposal of FL, differential privacy with SMC has been suggested as a privacy-preserving technique for the aggregation of independently trained neural networks in \cite{pathak2010multiparty}. Since then, this technique has been improved to return statistically indistinguishable results among participants while ensuring that such noise-added model parameters do not affect much on the accuracy of the global model in FL settings \cite{shokri2015privacy, aono2017privacy, geyer2017differentially, abadi2016deep, song2013stochastic, mcmahan2017learning}. As a consequence, adversaries cannot distinguish individual records in the FL training process and do not know whether or not a targeted client participating in the training; thus, preserving data privacy and protecting against the inference attacks. Generally, there are two types of employing differential privacy techniques for local nodes in FL settings: \textit{batch-level} and \textit{user-level} where random noise is added by measuring parameters' sensitivity from data points in a mini-batch and users themselves, respectively.

\subsubsection{Batch-level differential privacy approach}
Shokri and Shmatikov in \cite{shokri2015privacy} have proposed a communication efficient privacy-preserving SGD algorithm for deep learning in distributed settings in which local gradient parameters are asynchronously shared among participants with an option of adding noise to such updates for the differentially private protection of the individual model parameters. In this algorithm, participants can choose a fraction of parameters (randomly selected or following a strategy) to be updated at each round so that their local optimal can converge faster while being more accurate. In order to integrate differential privacy technique into the algorithm, the $\varepsilon$ total privacy budget parameter and the sensitivity of gradient $\Delta f_i$ for each parameter $f_i$ are taken into account to control the trade-off between the differential privacy protection and the model accuracy. \textit{Laplacian mechanism} is used to generate noise during both parameter selection and exchange processes based on the estimation of the $\Delta f_i$ sensitivity and the allocated $\varepsilon$ privacy budget. The proposed algorithm has experimented on \textit{MNIST} and \textit{SVHN} datasets showing the trade-off between strong differential privacy guarantees and high accuracy of the training model. However, with a large number of participants sharing a large fraction of gradients, the accuracy of the proposed algorithm with differential privacy is better than the standalone baseline. It is worth noting that in this algorithm, local gradients can be exchanged directly or via a central server, which can feasibly be implemented in the FL settings.

The authors in \cite{abadi2016deep} have proposed an SGD algorithm integrated with differential privacy performing over some batches (a group) of data samples. This algorithm estimates the gradient of the group by taking the average of the gradient loss of these batches and adds noise (generated by \textit{Gaussian mechanism}) to the group to protect the privacy. This algorithm is implemented to train on the \textit{MNIST} and \textit{CIFAR-10} datasets showing sensible results as it achieves only $1.3\%$ and $7\%$ less accurate compared to the non-differentially private conventional baseline algorithms on the same datasets, respectively. Similar to the mechanism proposed by Shokri and Shmatikov in \cite{shokri2015privacy}, the authors have proposed a mechanism to monitor the total privacy budget (i.e., privacy accounting) as accumulated privacy loss by observing privacy loss random variables. Based on the experiment, the authors also indicate that privacy loss is minimal for large group size (with a large number of datasets).

\subsubsection{User-level differential privacy approach}
Geyer \textit{et al.} in \cite{geyer2017differentially} have developed another method to implement differential privacy for federated optimisation in FL settings that conceals the participation of a user in a training task; as a result, the whole local training dataset of the user is protected against differential attacks. This approach is different from the \textit{batch-level} one, which aims at protecting a single data point in a training task. The proposed method utilises a similar concept of privacy accounting from \cite{abadi2016deep} that allows a coordination server to monitor the accumulated privacy budget by observing the moment accountant and privacy loss proposed in \cite{abadi2016deep}. The training process is halted once the accumulated privacy budget reaches a pre-defined threshold, implying that the privacy guarantee is no further tolerated. The Gaussian mechanism is also used to generate random noise which is then added to distort the sum of gradients updates to protect the whole training data. The proposed method has been experimented on \textit{MNIST} dataset, and the results show that with a sufficiently large number of participants (e.g., about 10,000 clients), the accuracy of the FL trained model almost achieves as high as the non-differential-privacy baseline while a certain level of privacy guarantee over the local training data still holds.

Similarly, McMahan \textit{et al.} in \cite{mcmahan2017learning} have leveraged the privacy accounting and moment privacy proposed in \cite{abadi2016deep} to integrate \textit{user-level} differential privacy into a federated averaging mechanism previously proposed in \cite{mcmahan2016federated} in order to protect local model parameters sharing with a coordination server. The proposed mechanism is a noise-added version of the federated averaging algorithm in FL which was deployed to train deep recurrent models like Long Short-Term Memory (LSTM) recurrent neural networks (RNNs). They have implemented the mechanism to train the LSTM RNNs tuned for language modelling in a mobile keyboard. The experimental results indicate that the integration of differential privacy only causes a minor effect on predictive accuracy; however, it could induce a qualitative effect on word predictions and tends to bias the model away from uncommon words. This potential bias in the mechanism calls for further research on adaptive tuning mechanisms for the clipping and noise in order to balance between utility and privacy in FL. Bhowmick \textit{et al.} in \cite{bhowmick2018protection} and Sun \textit{et al.} in \cite{sun2020ldp} have also proposed similar \textit{user-level} differential privacy in FL settings with some improvements such as employing a better estimation on total privacy budget (in \cite{bhowmick2018protection}), and adding a \textit{splitting \& shuffling} mechanism for local model parameters before sending to a coordination server (in \cite{bhowmick2018protection}).

As aforementioned, Hitaj \textit{et al.} have successfully carried out inference attacks at the client-side based on GAN \cite{hitaj2017deep}. In this paper, they have also shown that an FL training task with differential privacy employed at \textit{batch-level} is still susceptible to the attacks; however, the \textit{user-level} differential privacy approach could protect against such attacks.

\section{GDPR-Compliance in Centralised Federated Learning Systems}
FL emerges a new approach to tackle data privacy challenges in ML-based applications by decoupling of data storage and processing (i.e., local model training) at end-users' devices (i.e., local nodes) and the aggregation of a global ML model at a service provider's server (i.e., a coordination server). The privacy-preservation advantage of FL compared to the traditional centralised ML approaches is undeniable: It enables to train an ML model whilst retaining personal training data on end-users' devices. Only locally trained model parameters, which contain the essential amount of information required to update the global model, are shared with a coordination server. Nevertheless, such model parameters still enclose some sensitive features that can be exploited to reconstruct or to infer related personal information as depicted in \textit{Section 4}. Subsequently, an FL system still retains within the GDPR and is liable for complying with obligatory requirements. This section closely examines whether a GDPR requirement should be complied or inapplicable and should be waived in FL settings. Unsolved challenges on fully complying with the GDPR are also determined and discussed.

\subsection{Roles and obligations}
The GDPR differentiates three participant roles, namely Data Subject, Data Controller and Data Processor, and designates associated obligations for these roles under the EU data protection law. Data Controllers are subject to comply with the GDPR by determining the purposes for which, and the method in which, personal data is processed by Data Processors - who will be responsible for processing the data on behalf of Data Controllers. Furthermore, Data Controllers should take appropriate measures to provide Data Subjects with information related not only to how data is shared but also to how data is processed in the manner ensuring security and privacy of personal data. The GDPR also clearly specifies rights of Data Subjects, giving data owners the rights to inspect information about how the personal data is being processed (e.g., Right to be informed and Right of access) as well as to fully control the data (e.g., Right of rectification and erasure, and Right to restriction of processing).

As depicted in Table \ref{tb1}, in FL settings, personal data is regarded as local model parameters, not the original data samples as in traditional cloud-based ML systems. A service provider, who implements an FL system, is Data Controller and Data Processor combined as the service provider \textit{(i)} dictates end-users (i.e., Data Subject) to train an ML model using their local training data and to share such locally trained model, \textit{(ii)} processes the local model parameters sent from end-users (i.e., aggregates and updates the global model), and \textit{(ii)} disseminates the global models to all end-users and requests the end-users to update their local models. Furthermore, in centralised FL settings, a service provider can only share a global ML model, which can be considered as anonymous information, with third-parties as it does not possess any other personal data (e.g., original training data as in traditional ML systems). Therefore, Data Processors in FL settings are also the service providers, but not other players (i.e., third-parties). The processing mechanisms in FL are also uncomplicated compared to the traditional ones as they are only related to the aggregation of the local ML models as well as the update of the global ML model.

\begin{table*}[width=1.8\linewidth,cols=3,pos=h]
\centering
\caption{GDPR Roles in traditional centralised ML-based and centralised FL-based applications and services}
\label{tb1}
\begin{tabular*}{\tblwidth}{@{} LLL@{} }
\toprule
\textbf{GDPR Roles} & \textbf{Traditional ML-based services} & \textbf{Centralised FL-based services} \\
\midrule
Personal Data & Original training data & Local model parameters \\
Data Subject & End-users & End-users \\
Data Controller & Service Provider & Service Provider \\
Data Processor & Service Provider, Third-parties & Service Provider \\
\bottomrule
\end{tabular*}
\end{table*}

\subsection{The GDPR principles}
\begin{figure}[!htbp]
\centering
	\includegraphics[width=0.48\textwidth]{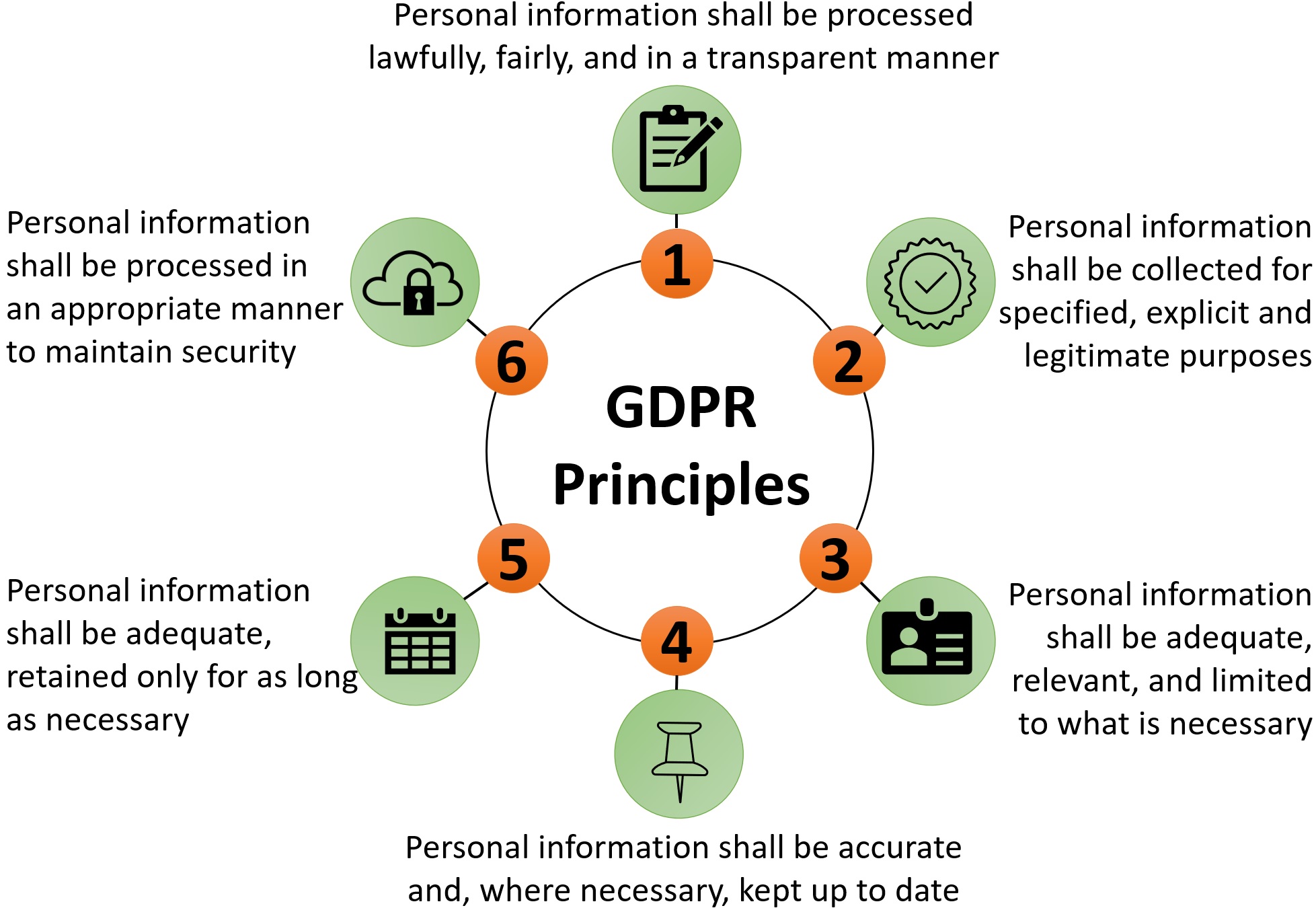}
	\caption{6-core principles in GDPR}
	\label{fig7}
\end{figure}

The GDPR defines 6-core principles as rational guidelines for service providers to manage personal data as illustrated in Fig. \ref{fig7} (The GDPR Articles 5-11). These principles are broadly similar to the principles in the Data Protection Act 1998 with the accountability that obligates Data Controllers to take responsibility for complying with the principles and implementing appropriate measures to demonstrate the compliance.

\subsubsection{Lawfulness, Fairness and Transparency}
According to the first principle, a service provider providing an FL application, as a Data Controller, must specify its legal basis in order to request end-users to participate in the FL training. There are total six legal bases required by the GDPR namely (1) Consent, (2) Contract, (3) Legal Obligation, (4) Vital Interest, (5) Public Task, and (6) Legitimate Interest (defined in Article 6 of the GDPR in detail). These lawful bases might need to come along with other separate conditions for lawfully processing some special category data including healthcare data, biometric data, racial and ethnic origin. Depending on specific purposes and context of the processing, the most appropriate one should be determined and documented before starting to process personal data.

To ensure privacy, an FL system is designed in a way that does not let the service provider (i.e., the coordination server) to directly access and obtain either original training data or locally trained ML models at end-users' devices. Instead, end-users, as participants in the FL system, will only send the results back to the coordination server when they are ready. An FL client-side application should offer several options for clients to participate in the training process proactively that allows a client to fully control the local training as well as of the sending/receiving ML model updates to/from a coordination server. Furthermore, FL systems only process data (i.e., local ML model parameters) for an explicit purpose (i.e., aggregates results and updates a global model), which is in ways that clients would reasonably expect whilst having minimal privacy impact. For these reasons, either \textit{Consent} or \textit{Legitimate Interest} legal basis can be appropriate for an FL application\footnote{https://ico.org.uk/for-organisations/guide-to-data-protection/guide-to-the-general-data-protection-regulation-gdpr/lawful-basis-for-processing/}.

Regarding the \textit{Fairness} and \textit{Transparency} requirements, as AI/ML algorithms like deep learning are normally operated in a black-box fashion, it is limited of transparency of how certain decisions are made, as well as limited understanding of the bias in data samples and training process \cite{doshi2017towards, mehrabi2019survey, ananny2018seeing, murdoch2019interpretable}. An FL system is not an exception. Generally, if the training data is poorly collected or intentionally prejudicial and fed to an ML, including FL, system, the results apparently turn out to be biased. If the trained model is then utilised for an automated decision-making system, then it probably leads to discrimination and injustice. Furthermore, the nature of preventing service providers from accessing original training dataset as well as the inability to inspect individuals' locally trained ML model due to Secure Aggregation mechanism amplifies the lack of transparency and fairness in FL systems. As a result, an FL system finds it problematic to transparently execute the training operations as well as to ensure any automated decisions from the system are impartially performed. This, consequently, induces the impracticality for any FL systems and fails to fully comply with the GDPR requirements of fairness and transparency.

These unsolved challenges appoint much more research on transparency, interpret-ability and bias for AI/ML algorithms as well as demand the GDPR supervisory boards to relax the requirements on AI/ML including FL systems. Another promising solution to comply with this GDPR principle is to design a new type of ML models with associated algorithms that are inherently interpretable, which encourages responsible ML governance \cite{rudin2019stop, li2017deep, molnar2020interpretable, harder2020interpretable}.

\subsubsection{Purpose Limitation}
This purpose limitation principle can be interpreted that an FL service provider needs to clearly inform clients about the purpose of a global ML model training as well as how clients' local personal data and devices' computation are used to locally train a requested ML model provided by the service provider. The principle also states that the service provider can further process the data for other compatible purposes. In this respect, FL systems fully satisfy with the principles if sufficient privacy-preserving mechanisms such as Secure Aggregation and differential privacy are readily implemented into the systems. This is because locally trained ML models from clients are aggregated only for the global model updates and cannot be individually extracted and exploited (by the coordination server) for other purposes.

However, as described in \textit{Section 4.1}, a malicious service provider or Byzantine participants can inject a hidden back-door model for an unauthorised training purpose. Currently, there is no solution for model anomaly detection mechanism at server-side for this type of attack due to the use of secure aggregation in centralised FL; this, as a consequence, remains as an unsolved challenge for an FL system to fully comply with the GDPR.

\subsubsection{Data Minimisation}
The data minimisation principle in the GDPR necessitates a Data Controller (e.g., a service provider) to collect and process personal data that is adequate, limited, and relevant only to claimed purposes. In traditional centralised ML algorithms, this data minimisation requirement is a challenge as it is not always possible to envision what data and the minimal amount of data are necessary for training an ML model. In this regard, FL appears as a game-changer as an FL system does not need to collect and process original training data; instead, a service provider only needs to gather local ML models from participants for assembling the global model. Generally, with privacy-preserving techniques introduced in \textit{Section 4}, an FL system can assure that the coordination server obtains aggregated local model parameters from participants for global model updates only (i.e., the claimed purposes) while acquiring nothing about individual's contribution. The aggregation mechanism also assures that the global model itself contains no individual sensitive features that can be exploited by adversaries to extract or infer any personal information.

Similar to the purpose limitation principle, back-door attacks are feasibly carried out to inject an unauthorised purpose. In this scenario, local ML model parameters obtained from participants is no longer minimal for the original purpose but also another unauthorised sub-task. This injected sub-task might be exploited to expose the personal information of the participant, imposing an unsolved challenge for FL systems.

\subsubsection{Accuracy}
The purpose of this principle is to ensure that a Data Controller should keep personal data correctly, updated, and not misleading any matter of fact. In centralised FL settings, a coordination server does not store any individual locally trained ML model parameters except the aggregated results from a batch of participants, and the anonymised global ML model. This information is stored and processed (i.e., for updating global model) in its original form without any changes, and updated for every training round. For these reasons, FL systems automatically satisfy the GDPR accuracy principle.

\subsubsection{Storage Limitation}
Basically, this principle ensures that a Data Controller does not keep personal data for longer if the data is no longer needed for the claimed purposes. In this case, data should be erased or anonymised. There is an exception for data retention only if the Data Controller keeps the data for public interest archiving, scientific or historical research, or statistical purposes.

Regarding the centralised FL settings, an FL system implementing Secure Aggregation does not store any individual ML model updates from participants except the global model - which can be assured to contain no individual sensitive features to be exploited for inference attacks. Even in the case of a malicious server holding aggregated contributions from many FL training rounds for further analytic (e.g., inference attacks), with secure aggregation and differential privacy integration, such aggregated information is protected and pseudo-anonymised. In other word, an FL system with appropriate privacy-preserving mechanisms can be fully compliant with the storage limitation principle.

\subsubsection{Integrity and Confidentiality (Security)}
This principle obligates Data Controllers to implement appropriate measures in place to effectively protect personal data. Thus, in order to comply with this principle, a centralised FL system requires to implement security and privacy mechanisms not only at a coordination server but also at end-users' devices as the FL system itself does not guarantee security and privacy.

Along with the privacy-preserving techniques such as Secure Aggregation, differential privacy, and Homomorphic Encryption designated for protecting local ML parameters, the FL client application installed at end-users' devices must be secure to prevent from unauthorised access, cyber-attack, or data breach directly from the devices or from the communications between the users' devices and a coordination server. This precondition is same as any other systems in which a variety of security and privacy techniques are readily integrated into FL applications, as well as secure communications protocols such as IPSec, SSL/TLS and HTTPS to protect data in transit between clients and the server.

\subsection{Rights of Data Subject}
The GDPR requires Data Controllers to provide the following rights for Data Subjects if capable (The GDPR Articles 12-23): (1) Right to be informed, (2) Right of access, (3) Right to rectification, (4) Right to erasure (Right to be forgotten), (5) Right to restrict processing, (6) Right to data portability, (7) Right to object, and (8) Rights in relation to automated decision making and profiling.

\subsubsection{Right to be informed} The challenge to provide this right to Data Subjects is that the GDPR demands the Data Controller to concisely, intelligibly, and specifically specify what and how the local ML model is used in the FL training, along with expected outputs of the mechanism\footnote{https://ico.org.uk/for-organisations/guide-to-data-protection/guide-to-the-general-data-protection-regulation-gdpr/individual-rights/right-to-be-informed}. Same as many complex ML mechanisms, FL is as a black-box model; thus, it cannot be precisely interpreted of how it works and predicting the outcomes. The GDPR supervisory board recognises the challenges and relaxes the requirement for AI/ML mechanisms by accepting a general explanation as an indication of how and what personal data is going to be processed. As a result, for an FL system, the right to be informed is achieved as \textit{privacy information} including purposes for processing local ML model (i.e., to build a global ML model), retention periods (i.e., no longer in use after each training round), and who it will be shared with (only the coordination server) can be determined as in Terms and Conditions when a client accepts to participate in an FL system.

\subsubsection{Rights in relation to automated decision making and profiling} A Data Subject is assumed to have the right "not to be subject to a decision based solely on automated processing, including profiling" - Article 22(1), the GDPR. Therefore, an FL client, as a Data Subject, has the right to receive meaningful information and explanation about whether the result of the processing (i.e., a global ML model) used in an automated decision-making system will produce legal effects concerning the client or similarly significantly affects the client. 
Unfortunately, due to the black-box operation model and the limitation of the transparency in ML, including FL, training process, results (e.g., a global ML model in FL) are generally generated without any proper explanation \cite{wachter2017right}. Thus, it is infeasible to predict whether outcomes of an ML model might affect the \textit{legal status} or \textit{legal rights} of the Data Subject, or negatively impact on its circumstances, behaviour or choices. Consequently, any FL system fails to comply with the GDPR requirements of the data subject's right in control of automated decision making. Fortunately, this requirement can be neglected if a Data Controller explicitly mentions the lack of automated decision making and profiling right when asking for Data Subject's consent to process personal data.

\subsubsection{Other Rights} The nature of decoupling between data storage and processing at client-side and global ML model aggregation at server-side in centralised FL leads to the unnecessity of providing the (2) Right of access, (3) Right to rectification, (4) Right to erasure, (5) Right to restrict processing, (6) Right to data portability, and (7) Right to object. For instance, regarding the "Right to erasure", if a user requests to delete its data (i.e., local ML model parameters sent to an FL server), literally, the only way to fulfil the user's request is to thoroughly re-train the global model without using user's data from the round that the user first participates \cite{ginart2019making}. This is unnecessary and impractical in FL settings as only local ML model parameters (possibly privacy guarantee-strengthened with differential privacy) in aggregated encrypted forms (by using Secure Aggregation and other advanced cryptography techniques) are shared with a coordination server. Consequently, it is worthless for a Data Subject to have control over its local ML model as \textit{(i)} the model parameters are protected by privacy-preserving techniques from inference attacks; \textit{(ii)} the server is unable to separate the user's data from the others, the server also does not store the model once it is aggregated to update the global model; and \textit{(iii)} the global model is wholly anonymised and cannot be exploited to extract or infer any individual information.

\subsection{GDPR-compliance Investigation and Demonstration}
The GDPR establishes supervisory authorities in each member state which are independent public authorities called Data Protection Authorities (DPAs). DPAs are responsible for supervising and inspecting whether a Data Controller is compliant with the data protection regulations whilst the Data Controller is responsible for demonstrating the compliance. The questions are judiciously raised: How can an FL system be investigated and validated by DPAs, and how can it demonstrate the compliance?

\subsubsection{DPA's supervisory competence}
\begin{figure}[!htbp]
\centering
	\includegraphics[width=0.48\textwidth]{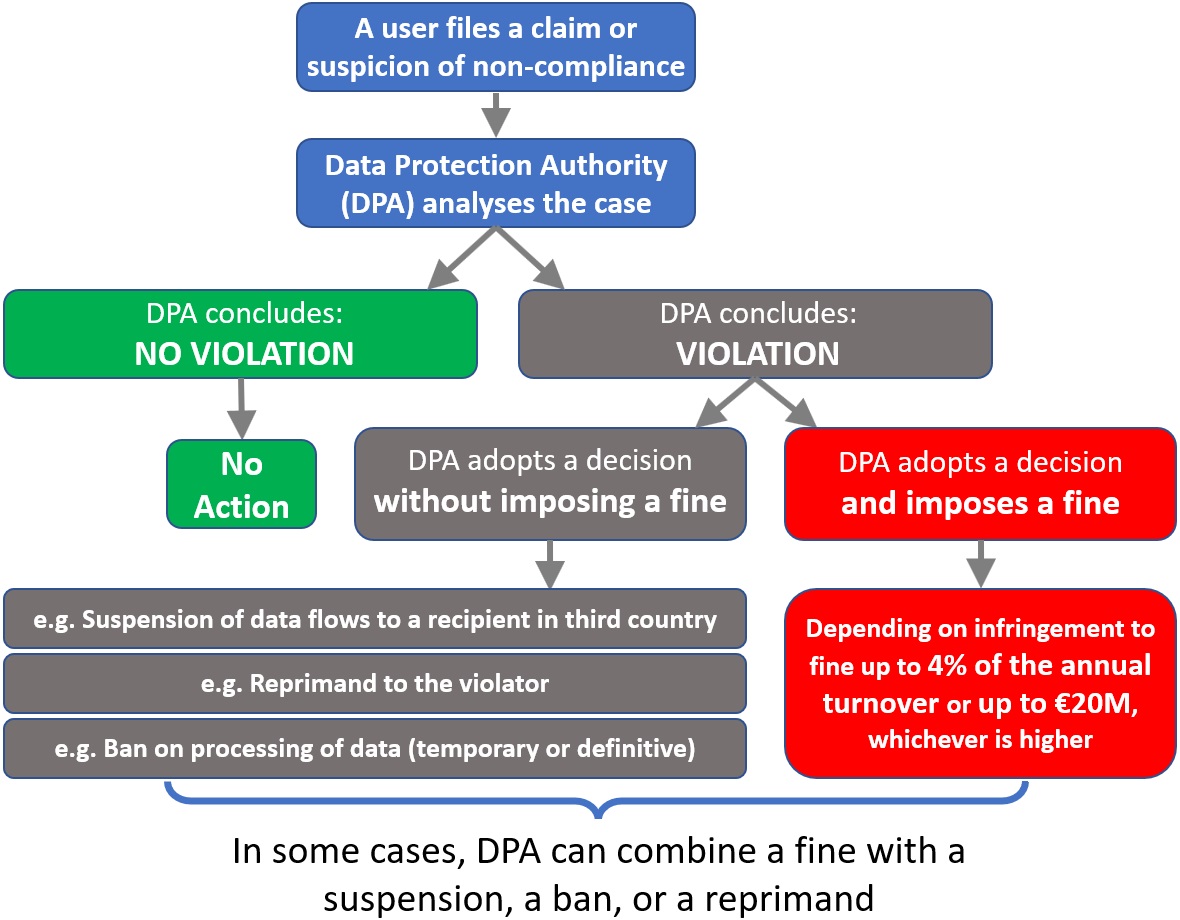}
	\caption{Workflow of the GDPR-compliance inspection and punishment procedure}
	\label{fig8}
\end{figure}

As illustrated in Fig. \ref{fig8}, the investigation of non\-/compliance and decision of punishment are carried out by DPAs once there is a suspicion or a claim filed by a customer. The compliance inspection will conduct some analysis to see whether a suspicious organisation follows the legal requirement of Privacy\&Security-by-design approach and satisfies some standard assessments such as Data Protection Impact Assessment (DPIA) and Privacy Impact Assessment (PIA), which are essential parts of the GDPR accountability obligations.

The GDPR establishes heavy punishment for non\-/compliance as failing to comply with the GDPR can be penalised by both financial fine (up to \EUR{}$20M$, or $4\%$ of global annual turnover, whichever is higher) and reprimand, ban or suspension of the violator's business (Fig. \ref{fig8}). A number of criteria specifically defined by the GDPR (Articles 77-84) are taken into account when determining the punishment such as the level of co-operation during the investigation, type of personal data, any previous infringement, and the nature, gravity, and duration of the current infringement. For instance, Facebook and Google were hit with a collective \$8.8 billion lawsuit (Facebook, 3.9 billion euro; Google, 3.7 billion euro) by Austrian privacy campaigner, Max Schrems, alleging violations of GDPR as it pertains to the opt-in/opt-out clauses. Specifically, the complaint alleges that the way these companies obtain user consent for privacy policies is an "all-or-nothing" choice, asking users to check a small box allowing them to access services. It is a clear violation of the GDPR's provisions per privacy experts and the EU/UK. A list of fines and notices (with non-compliance reasons) issued under the GDPR can be found on Wikipedia\footnote{https://en.wikipedia.org/wiki/GDPR\_fines\_and\_notices}

Normally, DPAs might require a variety of information with a detailed explanation from Data Controller to perform the analysis including documents of organisational and technical measures related to the implementation the GDPR requirements as well as independent DPIA and PIA reports frequently conducted by the Data Controller. DPAs may also require to be given access to data server infrastructure and management system including personal data that is being processed. In this respect, besides the legal basis such as consents from end-users, an FL service provider can only provide documentation of how FL-related mechanisms are implemented along with privacy-preserving technical measures such as secure aggregation, differential privacy, and homomorphic encryption. Other inquiries from DPAs such as direct access to the FL model training operations and inspection of individual local model parameters from a particular end-user are technically infeasible for any FL systems.

\subsubsection{Compliance Demonstration}
In order to build and demonstrate the GDPR compliance, AI/ML-based service providers should realise DPIA and PIA from the beginning of the project and document the processes accordingly which are designed to describe and clarify the whole data management processes along with the necessity and proportionality of these processes. Such assessments are important tools for accountability and essential to efficiently manage the data security and privacy risks, to demonstrate the compliance, as well as to determine the measure have been taken to address the risks. However, carrying out a DPIA or PIA is not mandatory for every data processing operation. It is only required when the operation is \textit{"likely to result in a high risk to the rights and freedoms of natural persons"} (Article 35(1)). The guideline for the criteria on the DPIA/PIA obligatory is described under Article 35(3), 35(4) which are adopted by DPAs to carry out such assessments.

In this respect, any FL service providers should perform the following steps for the DPIA/PIA to ensure the GDPR-compliance as well as to demonstrate the compliance once required by DPAs:

\begin{enumerate}
    \item A systematic description of data processing operations, associated purposes, along with clarification and justification of the operations. For instance, the operation of asking Data Subject's consent for local ML training and sending the ML model parameters to a coordination server should be documented in detail.
    
    \item An assessment of the necessity and proportionality of each operation, given its associated purposes. For instance, Secure Aggregation mechanism is necessary to implement whereas a differential privacy mechanism is proportionally required.
    
    \item An assessment of the data security and privacy risks that might be induced by each operation, along with the technical measures implemented to mitigate and manage the risks. For instance, in an FL system, the operation of sending local ML model parameters to a coordination server for global ML model update might be the target of inference attacks, thus, inducing privacy leakage. The measures called Secure Aggregation and Homomorphic Encryption mechanisms are implemented along with the technical report. Even though such privacy-preserving methods are implemented to strengthen FL systems, there exist some risks that can be exploited for illegitimate purposes such as model poisoning with back-door sub-tasks. These possible attacks, which lead to non-compliance with the GDPR, should be addressed.
\end{enumerate}

Foremost, same as any AI/ML-based system, an FL system is based on black-box complex ML models (e.g., deep learning and neural networks) with limited transparency, making it troublesome for both service providers and DPAs to comprehend and to inspect hidden operations taking place inside the system. Therefore, conducting DPIA/PIA on an FL system seems to be superficial, which requires much effort to discover breaches of the regulations, so as to avoid risky operations and to impose better privacy-preserving measures.
\section{Recap and Outlook}
AI/ML-based applications and services are high on the agenda in most sectors. However, the unregulated use or misuse of personal data is dramatically spreading, resulting in severe concerns of data privacy. A series of severe personal data breaches such as Facebook's Cambridge Analytica scandal, along with urgent mobile applications during the SARS-CoV2 pandemic for large-scale contact tracing and movement tracking \cite{ienca2020responsible} trigger worldwide attention respecting to a variety of privacy-related aspects including algorithm bias, ethics, implications of politic settings, and legal responsibility. This leads to a critical demand for effective privacy-preserving techniques, particularly for \textit{"data-hungry"} AI/ML-based systems, wherein FL is a prospective solution. The decoupling between local storage and processing at end-users devices and the aggregation of processing results at server-side in FL undoubtedly mitigate the risk of massive data breaches in a traditional centralised system while giving full control of personal data back to the users.

Although FL is in its infancy, we strongly believe that the collaborative computation with decentralised data storage as in FL systems has tremendous advantages to facilitate a variety of AI/ML-based applications without directly accessing end-users' data. Thus, FL systems are presumed to naturally comply with strict data protection legislation such as the GDPR. However, such FL systems still stay within the GDPR regulatory data protection framework as the local processing results sent to a server from end-users (e.g., locally trained ML model parameters) conceal some sensitive features that can be exploited to infer personal information of the end-users. Accordingly, FL systems are the target of some types of attack such as inference attacks and model poisoning, which could lead to infringements of the GDPR. Therefore, the systems must be strengthened by applicable privacy mechanisms such as SMC, differential privacy, and encrypted transfer learning methods \cite{salem2019utilizing}. We present a systematic summary of the threat models, possible attacks, and the privacy-preserving techniques in FL systems, along with the analysis of how these techniques can mitigate the risk of privacy leakages. Furthermore, insightful analysis of how an FL-system complies with the GDPR is also provided. Obligations and appropriate measures for a service provider to implement a GDPR-compliant FL system are examined in details following the rational guidelines of the GDPR six principles.

As FL is in the early stage, a fruitful area of multi\-/disciplinary research is commenced in order to flourish the technology and to comply with the GDPR fully. Firstly, efficient cryptographic and privacy primitives for decentralised collaborative learning must be further developed, particularly for counteracting model poisoning and inference attacks. Furthermore, as these privacy-preserving techniques such as SMC impose non-trivial performance overheads, further effort on how to efficiently utilise such techniques on FL applications are required. Secondly, research on transparency, interpret-ability and algorithm fairness in FL systems should be profoundly carried out. Even though a substantial amount of research has been conducted in centralised AI/ML settings, there is still an open question whether these approaches could be employed and how to sensibly adapt them to the decentralised settings where training data is highly skewed \textit{non-IID} and unevenly distributed across sources. The sampling constraints should be investigated to see how much extend they affect and how to mitigate the bias of the global training model. For instance, the agnostic FL framework introduced in \cite{mohri2019agnostic} naturally yields \textit{good-intent fairness} as it modelled the target distribution as an unknown mixture of the distributions instead of the uniform distribution in typical FL training algorithms. This agnostic FL framework, as a result, can control for bias in the training objective. Thirdly, it requires more research on interpretable and unbiased ML models and algorithms that can be employed over encrypted settings to well consolidate with advanced encryption schemes in FL systems. Besides, the trade-offs between privacy utility, accuracy, interpretability, and fairness in an FL framework need to be thoroughly explored.

If these requisites are successfully settled, it will assure to inaugurate responsible, auditable and trustworthy FL systems; as a result, complying with stringent requirements of the GDPR whilst bolstering the universal recognition of the secure decentralised collaborative learning solutions by both end-users and policymakers, including the GDPR supervisory authority.

\section*{Acknowledgement}
This research was supported by the HNA Research Centre for Future Data Ecosystems at Imperial College London and the Innovative Medicines Initiative 2 IDEA-FAST project under grant agreement No 853981.

\bibliographystyle{cas-model2-names}

\bibliography{refs}

\newpage

\bio{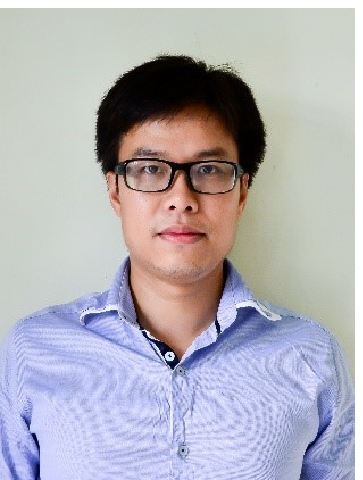}
Dr. Nguyen B.Truong is currently a Research Associate at Data Science Institute, Imperial College London, United Kingdom. He received his Ph.D, MSc, and BSc degrees from Liverpool John Moores University, United Kingdom, Pohang University of Science and Technology, Korea, and Hanoi University of Science and Technology, Vietnam in 2018, 2013, and 2008, respectively. He was a Software Engineer at DASAN Networks, a leading company on Networking Products and Services in South Korea in 2012-2015. His research interest is including, but not limited to, Data Privacy, Security, and Trust, Personal Data Management, Distributed Systems, and Blockchain.
\endbio

\bio{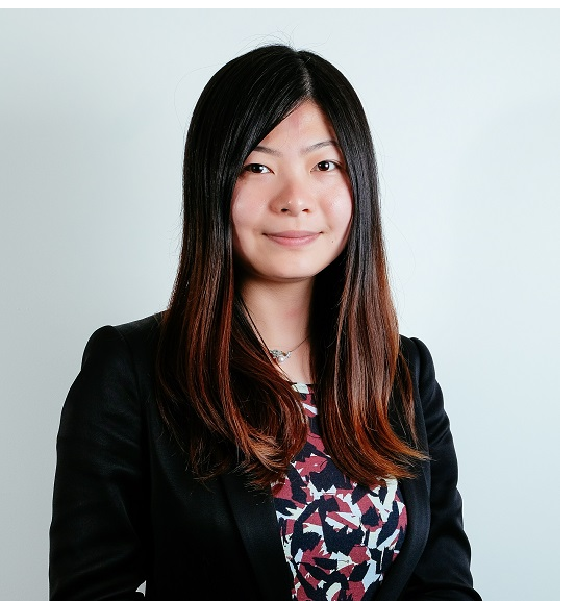}
Dr. Kai Sun is the Operation Manager of the Data Science Institute at Imperial College London. She received the MSc degree and the Ph.D degree in Computing from Imperial College London, in 2010 and 2014, respectively. From 2014 to 2017, she was a Research Associate at the Data Science Institute at Imperial College London, working on EU IMI projects including U-BIOPRED and eTRIKS, responsible for translational data management and analysis. She was the manager of the HNA Centre of Future Data Ecosystem in 2017-2018. Her research interests include translational research management, network analysis and decentralised systems.
\endbio

\bio{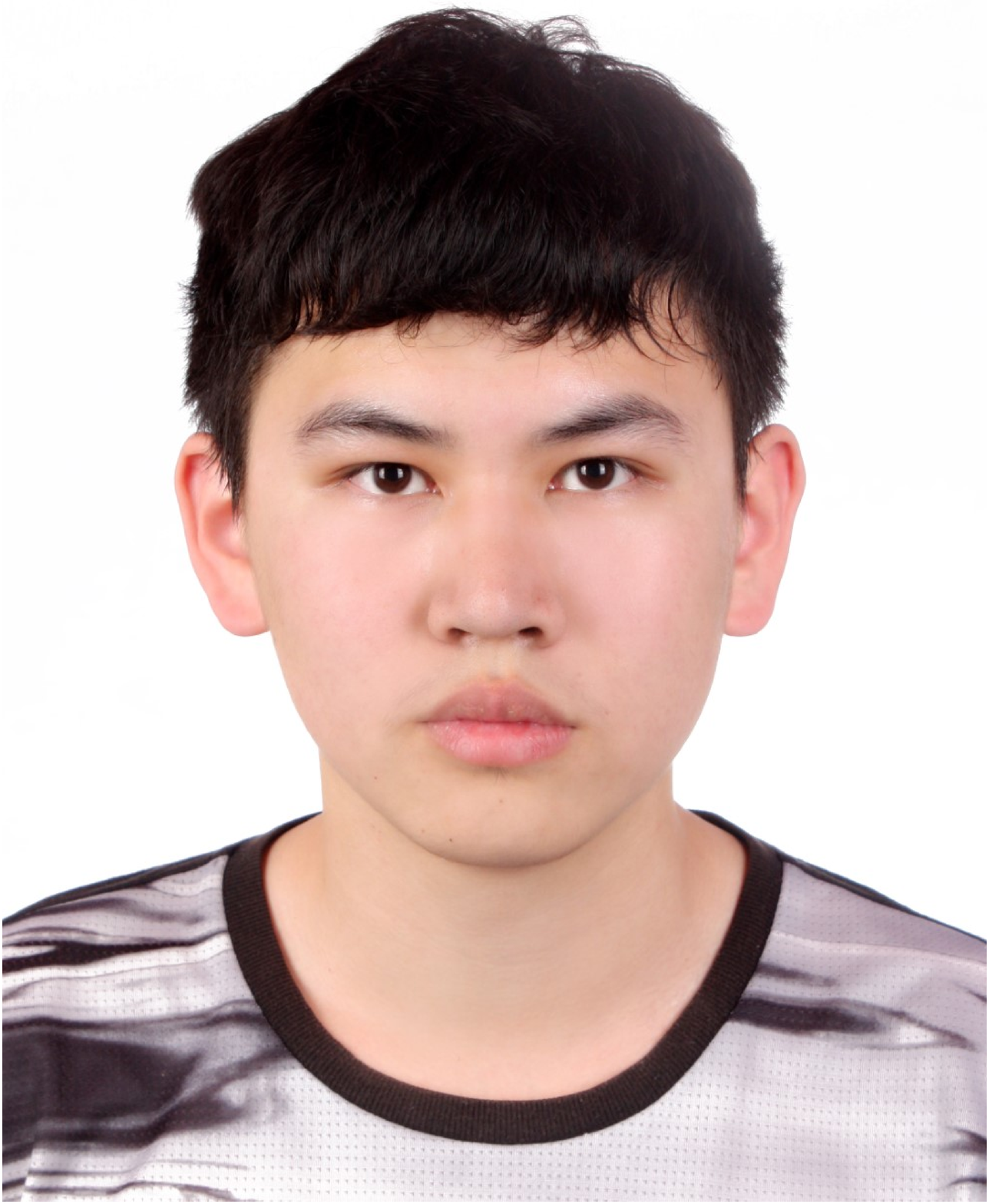}
Mr. Siyao Wang is a PhD student of the Data Science Institute at Imperial College London.
He received the BSc degree in Computer Science and Technology from the University of Chinese Academy of Sciences in 2018. He received the MRes degree in Medical Robotics and Image-Guided Intervention from Imperial College London in 2019. His research interests include machine learning, deep learning, computer vision and artificial intelligence applications in healthcare.
\endbio

\bio{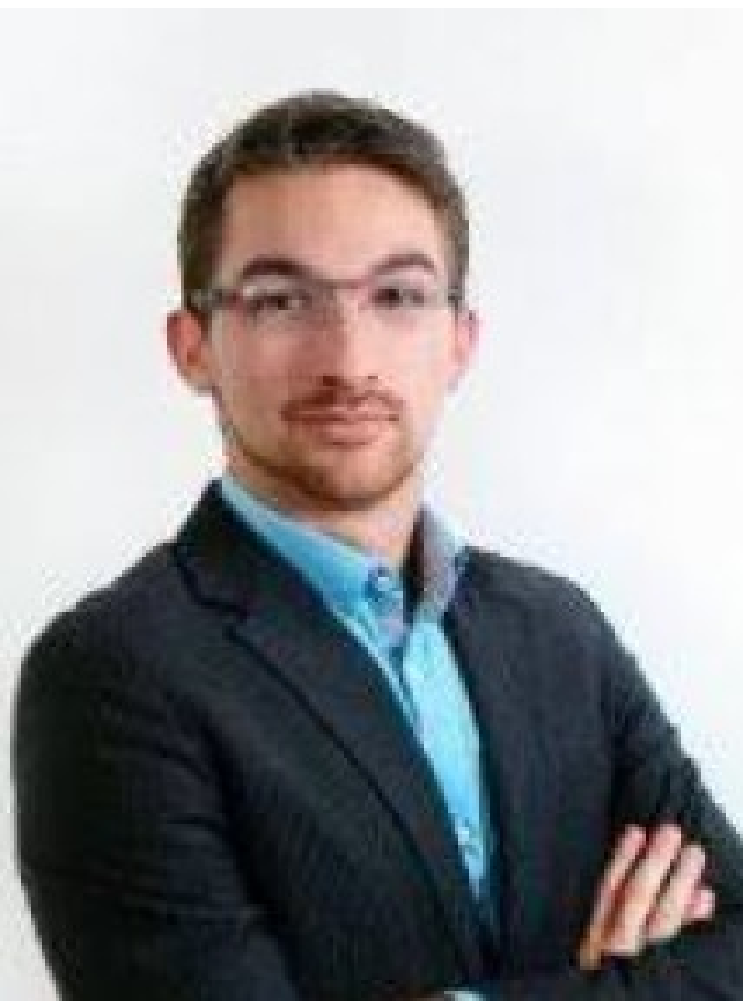}
Mr. Florian Guitton received a BSc in Software Engineering from Epitech (France) in 2011 and a MSc in Advanced Computing from the University of Kent (United Kingdom) in 2012. In 2012 he joined the Discovery Sciences Group at Imperial College London where he became Research Assistant working on iHealth, eTRIKS and IDEA-FAST EU programs. He is currently a PhD candidate at Data Science Institute, Imperial College London working on distributed data collection and analysis pipeline in mixed-security environments with the angle of optimising user facing experiences.
\endbio

\bio{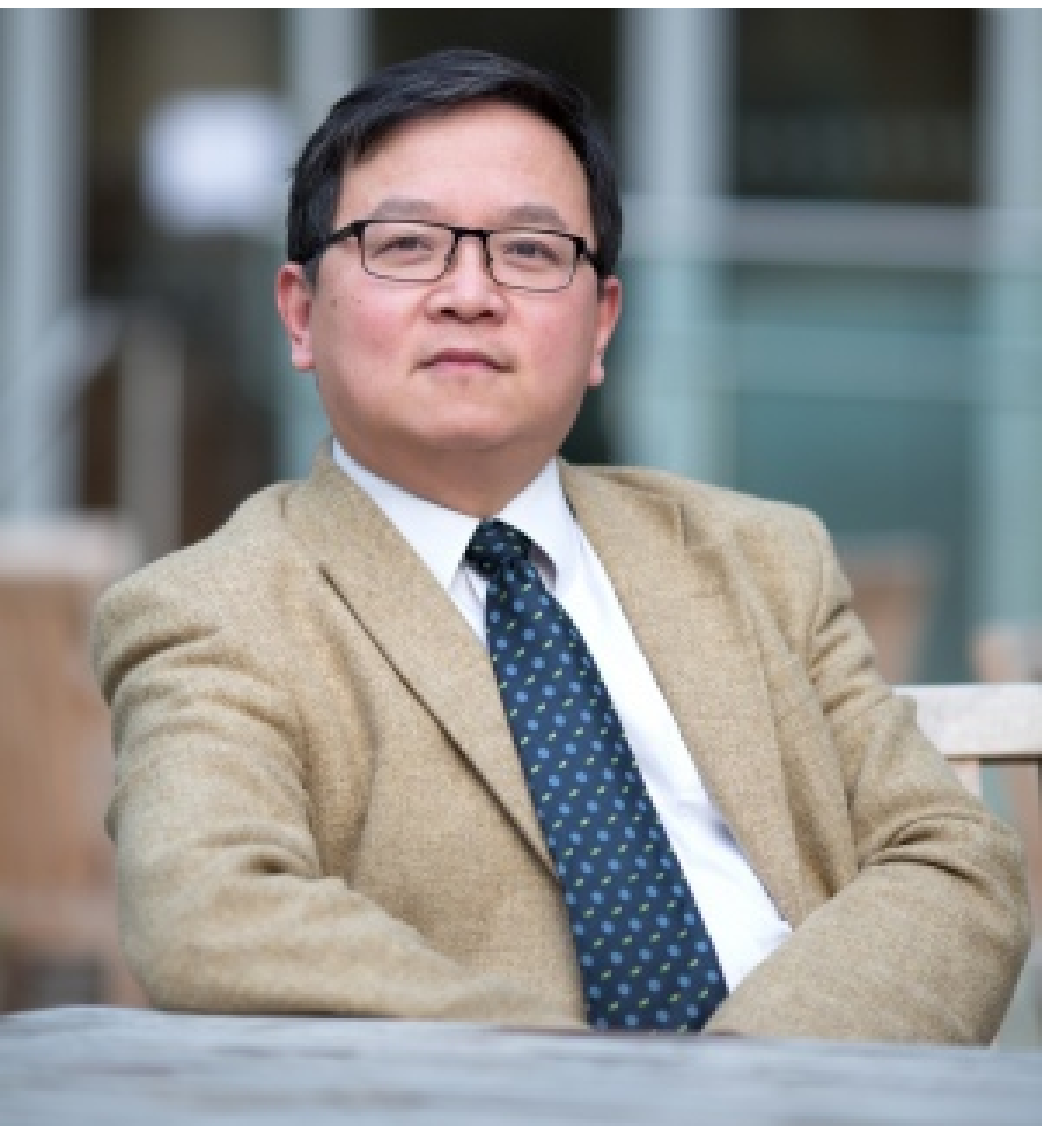}
Dr. Yike Guo (FREng, MAE) is the director of the Data Science Institute at Imperial College London and the Vice-President (Research and Development) of Hong Kong Baptist University. He received the BSc degree in Computing Science from Tsinghua University, China, in 1985 and received the Ph.D in Computational Logic from Imperial College London in 1993. He is a Professor of Computing Science in the Department of Computing at Imperial College London since 2002. He is a fellow of the Royal Academy of Engineering and a member of the Academia Europaea. His research interests are in the areas of data mining for large-scale scientific applications including distributed data mining methods, machine learning and informatics systems.
\endbio

\end{document}